# Hydrated peridotite – basaltic melt interaction Part II: Fast assimilation of serpentinized mantle by basaltic magma

Anastassia Y. BORISOVA[1,2*], Nail R. ZAGRTDENOV[1], Michael J. TOPLIS[3], Georges CEULENEER[1], Oleg G. SAFONOV[4, 2, 7], Gleb S. POKROVSKI[1], Klaus Peter JOCHUM[5], Brigitte STOLL[5], Ulrike WEIS[5], Svyatoslav SHCHEKA[6], Andrew Y. BYCHKOV[2]

[1] *Géosciences Environnement Toulouse, GET, Université de Toulouse, CNRS, IRD, UPS, France, 14 Avenue E. Belin, 31400 Toulouse, France*

[2] *Geological Department, Lomonosov Moscow State University, Vorobievy Gory, 119899, Moscow, Russia*

[3] *Institut de Recherche en Astrophysique et Planétologie (IRAP) UPS OMP – CNRS - CNES 14 Avenue E. Belin, 31400 Toulouse, France*

[4] *Korzhinskii Institute of Experimental Mineralogy, 142432, Chernogolovka, Moscow region, Russia*

*5 Climate Geochemistry Department, Max Planck Institute for Chemistry, P.O. Box 3060, D-55020 Mainz, Germany*

[6] *Bavarian Research Institute of Experimental Geochemistry and Geophysics (BGI), University of Bayreuth, 95440 Bayreuth, Germany*

[7] *Department of Geology, University of Johannesburg, Johannesburg, South Africa*

*Corresponding author: E-mail: anastassia.borisova@get.omp.eu;

*Corresponding address: Géosciences Environnement Toulouse UMR 5563, Observatoire Midi Pyrénées, 14 Avenue E. Belin, 31400 Toulouse, France; Tel: +33(0)5 61 54 26 31; Fax: +33(0)5 61 33 25 60

## Abstract

The most abundant terrestrial lavas, mid-ocean ridge and ocean island basalt (MORB and OIB), are commonly considered to be derived from a depleted MORB-mantle component (DMM) and more specific, variably enriched mantle plume sources. However, findings of oceanic lavas and mafic cumulates issued from melts, enriched in chlorine and having a radiogenic $^{87}Sr/^{86}Sr$ ratio, can be attributed to an interaction between the asthenosphere-derived melts and lithospheric peridotite variably hydrated due to penetration of hydrothermal water down to and below Moho level. To constrain mechanisms and rates responsible for the interaction, we report results of 15 experiments of reaction between serpentinite and tholeiitic basaltic melt at 0.2 - 1.0 GPa and 1250 - 1300°C. Results show that the reaction proceeds via a multi-stage mechanism: (i) transformation of serpentinite into Cr-rich spinel-bearing harzburgite ($Fo_{92-95}$ mol.%) containing pore fluid, (ii) partial melting and dissolution of the harzburgite assemblage with formation of interstitial hydrous melts (up to 57 – 60 wt% of $SiO_2$ contents at 0.5 GPa pressure), and (iii) final assimilation of the Cr-rich spinel-bearing harzburgite/dunite and formation of hybrid basaltic melts with 12 – 13 wt.% of MgO and elevated Cr (up to ~500 ppm) and Ni (up to ~200 ppm) contents. Assimilation of serpentinite by basaltic melt may occur under elevated melt/rock ratios (>2) and may lead to chromitite formation. We show that hybrid magmas produced by the progressive assimilation of serpentinized lithospheric mantle may be recognized by high Mg-numbers and high Cr and Ni contents of olivine and pyroxenes, an excess of $SiO_2$, $H_2O$ and halogens in the melts, and some unusual isotopic composition (e.g., radiogenic $^{87}Sr/^{86}Sr$, non-mantle $\delta^{18}O$ and low $^3He/^4He$). Our experiments provide evidence that MORB and high-Mg-Cr orthopyroxene-rich cumulates depleted in incompatible elements can be produced from common mid-ocean ridge basaltic melts modified by reaction with hydrated lithospheric peridotite. We established that the rate of assimilation of serpentinized peridotite is controlled by silica diffusion in the reacting hydrous basaltic melt. Our study challenges traditional interpretation of the variations in MORB and OIB chemical and isotopic composition in terms of deep mantle plume source heterogeneities or/and degrees of partial melting.

## Introduction

Mid-ocean ridge basalt (MORB) and ocean island basalts (OIB) are considered as products of decompression melting of several asthenospheric sources characterized by contrasted isotopic signatures (e.g., DMM, HIMU, EM-1 and EM-2; e.g., Zindler and Hart, 1986). This paradigm implies that the oceanic basalt composition is some kind of "carbon copy" of their deep mantle sources. However, the occurrence of basaltic glasses and high-Mg cumulates with radiogenic $^{87}Sr/^{86}Sr$ ratios along present-day mid-ocean ridges (e.g. Ross and Elthon, 1993, Nonnotte et al, 2005; van der Zwan et al., 2017) and in ophiolites, (Amri et al., 1996; Benoit et al., 1999; Clenet et al., 2010; Lange et al., 2013) suggest more complex mechanisms for the generation of oceanic magmas involving the assimilation of altered lithospheric material. Experimental and melt inclusion studies highlight the important role of basaltic melt-lithospheric rocks reactions on the chemical composition of both mantle rocks and silicate melts (e.g., Morgan & Liang, 2003; Kvassnes and Grove, 2008; France et al., 2010; Van den Bleeken et al., 2010; 2011; Borisova et al., 2012a; 2014; 2017). For example, Sr isotope diversity of basalts and mafic cumulates possibly related to assimilation of seawater-altered rocks is common in spreading environment (Michael and Cornell, 1998; Lange et al., 2013; van der Zwan et al., 2017). Magmas resulting from the mixing between "asthenospheric" tholeiites and hydrated "lithospheric" melts of depleted andesitic affinity were proposed as parental for some puzzling occurrences of orthopyroxene-rich primitive cumulates along mid-ocean ridges and in ophiolites (Benoit et al, 1999; Nonnotte et al., 2005). The formation of such hybrid or/and depleted magmas due to assimilation of serpentinized mantle peridotite leads to the formation of the Moho transition zone composed by dunites and chromitites (Borisova et al., 2012a; Zagrtdenov et al., 2018; Rospabé et al., 2019a,b).

Hydrothermal circulation in the oceanic lithosphere near and below the mantle/crust boundary ("petrologic Moho") is recorded in such hydrothermal rocks as mantle diopsidites hosted by peridotites in the shallow mantle section of ophiolites (Python et al., 2007; Akizawa & Arai, 2014) as well as in multiphase inclusions representing fluid phase, not a magma, in chromitite orebodies (Borisova et al., 2012a; Johan et al., 2017). Assimilation of seawater-derived brine or aqueous fluid by mafic rocks, magmas and melts, is recorded also at ultraslow-, slow-, and fast-spreading mid-ocean centers and preshield-stage Loihi Seamount lavas (Hawaii) (Michael & Schilling, 1989; Jambon et al., 1995; Michael & Cornell, 1998; Kent et al., 1999; Dixon & Clague, 2001; Dixon et al., 2008; Klein et al., 2019). Seawater-derived component assimilation, possibly through interaction with serpentine, happens at slow-spreading centers or at small oceanic islands (Simons et al., 2002; Dixon et al., 2008). It is also reported that basaltic glasses, melt inclusions and the serpentinite-hosted gabbroic veins in the environment of the slow-spreading ridges are affected by assimilation of hydrothermally altered lithosphere enriched in such volatile elements as Cl, $H_2O$, and atmospheric Ne, Ar isotopes (Stroncik & Niedermann, 2016; van der Zwan et al., 2017; Ciazela et al., 2017; 2018). The shallow depths of the melt- or magma-rock interactions extend to 10 to 13 km, suggesting not only crustal but also upper mantle source of such contamination (van der Zwan et al., 2017). Additionally, serpentinites play an important role during melt-rock interactions of upper mantle, which was thereafter exposed in the detachment faults (e.g., Bach et al. 2004; Sauter et al., 2013; Ciazela et al., 2017), where upper mantle serpentinites may be abundant in the crust of slow and (mostly) ultraslow-spreading ridges. Indeed, hydrated peridotites, in particular, upper mantle serpentinites are high-Mg rocks variously enriched in refractory elements (Cr, Ni) and fluid-mobile elements (e.g., halogens, H, B, O, He, Ar, As, S, Sb, Sr and Pb) due to interaction between peridotite and seawater-derived low-to-moderate-temperature hydrothermal fluids (e.g., Guillot et al., 2001; Früh-Green et al., 2004; Bonifacie et al., 2008; Deschamps et al., 2010; Evans et al., 2013; Guillot & Hattori, 2013; Kendrick et al., 2013). These rocks crop out preferentially below the mantle-crust transition zone (Evans et al., 2013) which is locally exposed to the seafloor at mid ocean ridges (e.g., Bach et al., 2004) and may be sampled as upper mantle xenoliths by lavas of such oceanic islands as Canary Islands (Neumann et al., 2015) at possible reaction depths of up to 1.0 GPa pressure.

Although the interaction between dry tholeiitic basalt and anhydrous harzburgite has been investigated experimentally by Fisk (1986), Kelemen et al. (1990), Morgan and Liang (2003), and Van den Bleeken et al. (2010; 2011), the rates and mechanisms of assimilation of the hydrated mantle lithosphere (or serpentinite rocks) by the basaltic magmas and the chemical impact of the hydrated mantle on the composition of oceanic basalts are still unconstrained. Such constraints are important to understand the evolution of the oceanic lithosphere and to reconsider some inferences of geochemists on the nature and composition of mantle sources (Nonnotte et al., 2005; Kendrick et al., 2017). Our experiments were designed to study the reaction processes between moderately differentiated basaltic melt and serpentinite with high melt/rock ratio at pressures of 0.2 to 1.0 GPa. Our new data contribute to constrain the rate, mechanism and compositional impact of the assimilation of the serpentinized lithospheric mantle by the basaltic magma.

## Materials and Methods

### Starting Materials and Analytical Methods

The mid-ocean ridge basalt used in the experiments is a typical moderately differentiated (8.2 wt.% of MgO) glassy tholeiitic basalt (sample 3786/3) from Knipovich ridge of the Mid Atlantic Ridge dredged during the 38$^{th}$ Research Vessel Adademic Mstislav Keldysh expedition (Sushchevskaya et al., 2000). The serpentinite used as starting material is a homogeneous rock composed by antigorite with accessory Fe-rich oxides, devoid of relics of primary mantle silicates, sampled in Zildat, Ladakh, northwest Himalaya (e.g., Deschamps et al., 2010). The composition of the starting materials is given in **Table 1**. The concentrations of elements in both rocks were measured at the SARM (Service d'Analyse des Roches et des Minéraux, Centre de Recherches Pétrographiques et Géochimiques, Vandoeuvre lès Nancy, France). The rock $H_2O$ concentrations were measured using Karl Fischer titration. Major and trace elements were measured using inductively coupled plasma optical emission spectroscopy (ICP–OES) and ICP–MS, using a method developed at the SARM (Carignan et al., 2001) employing an ICP–OES IRIS Advantage ERS from Thermo Scientific and an ICP–MS x7 from Thermo Scientific.

For the hybrid runs, the serpentinite has been prepared as doubly polished ~1000 µm-thick section, thereafter cut to 2.7 mm-diameter cylinders by a core drill machine. The MORB glass has been crushed to powder (<100 µm glass size). Additionally, for the mixed runs at 0.2 GPa pressure, the serpentinite sample has been crushed to powder (<100 µm glass size).

**Experimental Strategy and Method**

The experiments in the system containing basalt and serpentinite were performed at 0.2 to 1.0 GPa and 1250 to 1300°C. Although in modern oceanic settings, temperatures of 1050°C are sufficient to initiate reaction of hydrated peridotite with basaltic magma at 0.2 GPa (Borisova et al., 2012a), we chose higher temperatures for the experiments as higher temperatures substantially increase reaction rates of the serpentinite-basaltic melt interactions and ensure conditions corresponding to complete melting of the basalt, consistent with the majority of existing models of basaltic melt extraction from the mantle (Fisk, 1986; Hirschmann et al., 1998; Ulmer, 2001; Morgan and Liang, 2003).

Fifteen experimental runs were performed at 0.2 to 1.0 GPa with similar serpentinite to basalt ratios (with 12 to 28 wt% of serpentinite in the mixture), the similar initial bulk water content and different run duration (**Fig. 1, Table 2**). Two series of experiments were performed: (a) ***hybrid*** series with serpentinite cylinder and basaltic powder at 0.2 and 1.0 GPa and (b) ***mixed*** series composed of intimately and well-mixed serpentinite and basaltic powders uniquely at 0.2 GPa pressure. The experimental design of the hybrid runs included a serpentinite cylinder in the upper part and dry or wet MORB glass powder in the lower part of the $Au_{80}Pd_{20}$ (exceptionally, in one run of the Pt) capsule. The mixed runs were performed with mixtures of basaltic powders and 20 wt% of serpentinite powder in the bulk mixture (**Fig. 1, Table 2**). The starting components were weighed before runs. The distilled water$^{MQ}$ was mixed with the MORB powder in the experiments with additional water uniquely at 0.5 GPa pressure. The experimental runs were performed at temperature of 1250 – 1300°C. Because the run duration was shorter than 48 hours which is necessary to reach oxygen fugacity equilibrium with a piston-cylinder double-capsule techniques of mineral buffer (Matjuschkin et al., 2015), the redox conditions in our kinetic experiments were controlled by the initial $Fe^{2+}/Fe^{3+}$ ratios in the MORB glass and serpentinite, although more oxidized conditions were established during the runs due to the presence of water and partial $H_2$ loss to $Al_2O_3$ pressure media. The $Fe^{2+}/Fe^{3+}$ ratios in several quenched products were estimated using XANES (X-ray absorption near edge structure) at European Synchrotron Radiation Facility (ESRF) in Grenoble (France). Oxygen fugacity relative to the quartz-fayalite-magnetite buffer (QFM) was calculated based on the obtained ferric iron ($Fe^{III}$) mole fraction ($X_{FeIII}$) in the analyzed samples (**Table 2**) using of the $FeO^{total}$ contents in the predominant basaltic melt and model of Borisov et al. (2018). The redox conditions in the shortest runs (<4 h) were estimated from the olivine-chromite assemblages using equations of Ballhaus et al. (1991) and from mole fraction of ferric iron $Fe^{III}$ obtained by XANES at corresponding temperatures to range from QFM(+1.5) (for P10) to QFM(+3.6) (for P1) (**Fig. 1, Table 2**).

Two piston-cylinder systems were used in our experiments. Experiments (P1 – P10) used the end-loaded Boyd-England piston-cylinder apparatus at the Korzhinskii Institute of Experimental Mineralogy, Chernogolovka, Russia. Standard talc-Pyrex cells 3/4 inch in diameter, equipped with tube graphite heaters and inserts made of MgO ceramics were used as pressure-transmitting medium. The pressure at elevated temperatures was calibrated against two reactions of brucite = periclase + $H_2O$ and albite = jadeite + quartz equilibria. A pressure correction (12%) was introduced for the friction between the cell and hard-alloy vessel. To minimize the friction, a Pb foil and molybdenum disulfide ($MoS_2$) lubricant were used. $Au_{80}Pd_{20}$ capsules with starting mixtures were mounted in the central parts of the cells. The temperature in the upper part of the capsules was controlled to be accurate to ±1°C using a MINITHERM controller via a $W_{95}Re_5/W_{80}Re_{20}$ thermocouple insulated by mullite and $Al_2O_3$ without pressure correction. For the Pyrex-bearing assemblies the sample was heated to 550 – 600°C at low confining pressure (0.15 – 0.2 GPa) for a few minutes in order to soften the Pyrex glass, subsequently both temperature and pressure were increased almost simultaneously up to the desired run conditions. The samples were maintained at run conditions during desired durations (**Table 2**). The experiments were quenched by switching off electricity. The quench rate was 100–300°C/min.

The experiments (P15 – P26) used the “Max Voggenreiter” end-loaded Boyd-England piston-cylinder apparatus at the Bavarian Research Institute of Experimental Geochemistry and Geophysics (BGI), Bayreuth, Germany. Talc cells 3/4 inch in diameter with Pyrex sleeves were used. A tapered graphite furnace was inserted in each cell. Alumina ($Al_2O_3$) spacers were used as pressure-transmitting medium. An $Au_{80}Pd_{20}$ capsule loaded with starting materials was set in the central part of the assembly. A 20% pressure correction was applied for the friction between the talc cell and pressure vessel. A molybdenum disulfide ($MoS_2$) lubricant was introduced to minimize the friction. The temperature in the upper part of the capsules was controlled by a EUROTHERM (2404) controller via either $W_3Re_{97}/W_{25}Re_{75}$ (type D) or $Pt_6Rh_{94}/Pt_{30}Rh_{70}$ (type B) thermocouple accurate to ±0.5°C. The sample was compressed to 0.5 GPa during a period of 20 minutes and then heated up to the run temperature (1300 °C) at a rate of 100 °C/min. The samples were maintained at run conditions during desired durations (**Table 2**). The experiments were quenched by switching off electricity. We have applied decompression during periods from 20 minutes to 2 hours. The rate of quenching to the ambient temperature was ~ 300°C/min.

Five experiments (P36, P37, SB1, SBbis3, and SBter1) were carried out at pressure of 0.2 to 0.5 GPa and a temperature of 1250 °C using an internally heated gas pressure vessel at the Korzhinskii Institute of Experimental Mineralogy, Chernogolovka, Russia. The pressure in the system was created by pure Ar gas. The system was heated by a furnace with two windings (to minimize the thermal gradient). The temperature was set and measured by a TRM-101 OVEN controller through two S-type ($Pt_{90}Rh_{10}$ vs $Pt_{100}$) thermocouples. The thermocouples were mounted at the top and close to the bottom of the run hot spot to monitor the temperature gradient. The duration of experiment was from 0.5 to 48 h (**Table 2**). The experiments were quenched by switching off the furnace. The pressure during the quench was maintained constant down to 550 °C, and then slowly released. The cooling rate from 1250 to 1000 °C was 167 °C/min, and then 90 °C/min down to 550 °C. After the runs, the capsule was mounted in epoxy, cut in two parts using a diamond saw, and then polished using SiC sand papers and diamond pastes.

To calculate phase proportions, we have used the PYTHON language code. At the first step, the initial MORB glass was introduced instead of $L_{bas}$ and $L_{int}$. On the second step, the basaltic and interstitial glasses ($L_{bas}$ and $L_{int}$, respectively) were distinguished as two phases.

**Scanning Electron Microscope (SEM) and Electron Microprobe analysis (EPMA)**

Major and minor element analyses of minerals and glasses and the experimental sample imaging were performed at the Géosciences Environnement Toulouse (GET, Toulouse, France) laboratory using a scanning electron microscope (SEM) JEOL JSM-6360 LV with energy-dispersive X-ray spectroscopy (EDS), coupled with the automatic analyzer of particles by the program "Esprit". The main experimental phases (oxides, silicates and glasses) in the samples have been identified by EDS microprobe technique at GET (Toulouse, France) (Borisova et al., 2012b). Major, and minor element compositions of the crystals and glasses were analyzed using CAMECA SX-Five microprobe at the Centre de Microcaractérisation Raimond Castaing (Toulouse, France). Electron beam of 15 kV accelerating voltage, and of 20 nA current was focused or defocused on the sample to analyze minerals or glasses, respectively. The following synthetic and natural standards were used for calibration: albite (Na), corundum (Al), wollastonite (Si, Ca), sanidine (K), pyrophanite (Mn, Ti), hematite (Fe), periclase (Mg), Ni metal (Ni), and $Cr_2O_3$ (Cr). Element and background counting times for most analyzed elements were 10 and 5 s, respectively, whereas, peak counting times were 120 s for Cr and 80 to 100 s for Ni. Detection limits for Cr and Ni were 70 ppm and 100 ppm, respectively. The silicate reference materials of Jarosewich et al. (1980) as well as MPI-DING glasses of ultramafic to mafic composition (GOR132-G, GOR128-G, KL2-G and ML3B-G of Jochum et al., 2006) were analyzed as unknown samples to additionally monitor the analysis accuracy. The silicate reference material analysis allowed to control precision for the major and minor (e.g., Cr, Ni in glasses) element analyses to be in the limit of the analytical uncertainty (related to the count statistics). The accuracy estimated on the reference glasses ranges from 0.5 to 3 % (1σ RSD = relative standard deviation), depending on the element contents in the reference glasses.

$H_2O$ contents of glasses were estimated based on the *in situ* electron probe analyses of the major element oxides. The 0.5 – 1.0 GPa glasses were analyzed following simplified "by difference method" without bracketing. The uncertainty of the water contents generally varies between 10 and 50 %. The 0.2 GPa glasses were analyzed by using bracketing mode (Borisova et al., 2020).

## X-ray Absorption Near Edge Structure (XANES) Spectroscopy

Iron redox state in selected quenched glasses was determined from Fe K-edge (~7.1 keV) XANES spectra acquired at the FAME beamline (Proux et al., 2005) of the European Synchrotron Radiation Facility (ESRF). The beamline optics incorporates a Si(220) monochromator with sagittal focusing allowing an energy resolution of ~0.5 eV at Fe K-edge and yielding a flux of >1012 photons/s and a beam spot of about 300×200 µm. XANES spectra were acquired in fluorescence mode in the right-angle geometry using a 30-element solid-state germanium detector (Canberra). Energy calibration was achieved using a Fe metal foil whose K-edge energy was set to 7.112 keV as the maximum of the spectrum first derivative. Iron-bearing oxides and silicates with different Fe redox and coordination environment, diluted by mixing with boron nitride to obtain Fe concentrations of a few wt%, were measured similarly to the glasses to serve as reference compounds.

Iron redox state in the glasses was determined by fitting the XANES pre-edge region (7.108-7.120 keV) according to the protocols developed in Muñoz et al. (2013) and using a background polynomial and two pseudo-Voight functions to determine the energy position of the pre-edge peak centroid, which is a direct function of the $Fe^{III}/Fe^{II}$ ratio both in crystalline and glass silicate samples (Wilke et al. 2001; 2005). Ferric iron ($Fe^{III}$) mole fraction ($X_{FeIII}$) in the starting basalt, and sample P3 was determined using the calibration established for basaltic glasses (Wilke et al., 2005), while sample P1 that showed a mixture of glass and crystals was processed using the calibration established for Fe minerals (Wilke et al., 2001). The results are reported in **Table 2**. The uncertainties of $X_{FeIII}$ determination for dominantly glassy samples are 0.05 in absolute value, while those for glass-crystal mixtures are typically 0.07 of the value.

## Laser Ablation Inductively Coupled Plasma Mass Spectrometry

Major and trace element concentrations were determined by LA-ICP-MS at the Max Planck Institute for Chemistry, Mainz, using a New Wave 213 nm Nd:YAG laser UP 213, which was combined with a sector-field ICP-MS Element2 (Thermo Scientific) (Jochum et al., 2007; 2014). Ablation took place in the New Wave Large Format Cell under He atmosphere. Spot analyses were performed in low mass resolution mode using crater sizes of 30 µm (lines of spots) and 8 µm (single spots), and a pulse repetition rate of 10 Hz at a fluence of about 6.5 $J/cm^2$ (lines of spots) and 8.3 $J/cm^2$ (single spots). Isotopes used for analysis are the following: $^{7}Li$, $^{23}Na$, $^{25}Mg$, $^{27}Al$, $^{29}Si$, $^{31}P$, $^{39}K$, $^{43}Ca$, $^{47}Ti$, $^{53}Cr$, $^{55}Mn$, $^{57}Fe$, $^{62}Ni$, $^{85}Rb$, $^{88}Sr$, $^{89}Y$, $^{90}Zr$, $^{93}Nb$, $^{137}Ba$, $^{139}La$, $^{140}Ce$, $^{141}Pr$, $^{146}Nd$, $^{147}Sm$, $^{151}Eu$, $^{157}Gd$, $^{159}Tb$, $^{163}Dy$, $^{165}Ho$, $^{167}Er$, $^{169}Tm$, $^{173}Yb$, $^{175}Lu$, $^{178}Hf$, $^{232}Th$ and $^{238}U$. Data reduction was performed by calculating the ion intensities of each isotope relative to the intensity of $^{43}Ca$. The NIST SRM 610 silicate glass (for trace and major elements except Mg, K, Fe) and the basaltic glass GSE-1G (Mg, K, Fe) were used for calibration. Major element concentrations were calculated to a total oxide content of 99 wt%. The repeatability (RSD) of the measurements is about 1 - 3% (30 µm measurements) and 5 - 10 % (8 µm). The detection limits (3 σ definition) for the 30 µm measurements vary between about 0.001 and 1 ppm (Jochum et al., 2007; 2014). They are about a factor of 4 higher for the 8 µm analyses. Measurement accuracy was tested with GSE-1G. The concentration values agree within about 5 % (30 µm) and 10 % (8 µm) of the reference values (GeoReM database) (Jochum et al., 2006).

## Results of Experiments on Basaltic Melt-Serpentinite Interaction

### Experimental Sample Description

The first group of experiments was conducted at 0.5 GPa and 1300 °C with duration up to 8 hours (**Tables 2 – 4, Table A1**). Products of the shortest run P1 (1 minute at the run temperature) which was considered as a zero-time experiment, the quenched basaltic glass zone and a zone replacing serpentinite (former serpentinite zone) are present. The former serpentinite zone contains fine-grained (5 to 10 µm in size) aggregates of olivine $Fo_{95}$, enstatite (Mg# = 95), chromite (Cr# = 89) and interstitial glass of basaltic andesite composition (**Fig. 1**). Chromite crystals (a few micrometers in size) are disseminated within this zone.

Samples P15 and P10 were kept for 0.5 and 2.6 hours, respectively, at run conditions. Distinct quenched melt zone and the former serpentinite zone are also present in this sample. The melt zone consists of hydrous basaltic glass. The former serpentinite zone shows 5-20 µm-size aggregate of forsteritic olivine $Fo_{93}$ and enstatite Mg# = 97 (P15) or accessory clinopyroxene Mg# = 81 (P10) (**Fig. 1**). It is associated with interstitial glass of basaltic andesite to andesitic composition and chromiferous magnetite (P15) or chromite (P10). Chromiferous magnetite and chromite are clustered in two large (n × 100 µm) areas (P15) or disseminated in the olivine-rich zone (P10).

Sample P18 provides an information about 5 hours-lasting basaltic melt-serpentinite interaction. The sample shows a hydrous basaltic zone and former serpentinite zone. The former serpentinite zone consists of two areas. The outer area of nearly 200 µm width contains olivine, interstitial basaltic glass and disseminated chromite (grain size of a few micrometers). The inner area consists of forsteritic olivine $Fo_{93}$, enstatite (Mg# = 97) and contains magnetite aggregate (~20 × 40 μm) (**Fig. 1**).

P36 is the longest (8 hours) experiment of the series at 0.5 GPa. Serpentinite is completely dissolved in the basaltic melt, and the run products are represented by homogeneous basaltic glass with 11.5 wt% MgO.

P20, P21, and P26 runs were performed with an additional water and run duration of 0.5, 2.5 and 5.0 h, respectively (**Tables 2, A1**). The run products are composed of hydrous basaltic zone and former serpentinite zone where forsteritic olivine $Fo_{91-94}$, enstatite (Mg# = 93 – 96) and chromite are associated with interstitial glass of basaltic andesite to andesite composition (Mg# = 51 – 67). Numerous bubbles in the basaltic glass reflect saturation with the aqueous fluid at the run conditions.

The second series of experiments has been performed at 1.0 GPa and 1300 °C. Experiment P3 with duration 2.5 hours shows basaltic glass zone and the former serpentinite zone where forsteritic olivine $Fo_{92}$, enstatite (Mg# = 96) and Cr-bearing magnetite are associated with interstitial glass of basaltic composition (**Fig. 1**). The 9 h-long run without additional water (P7) and 3 h-long run with additional water (P12) contain uniquely hydrous basaltic glasses with 13.0 wt% MgO.

Thus, at 0.5 - 1.0 GPa pressure range, the initial stage of basaltic melt-serpentinite reaction generates two contrasting zones: an olivine-rich zone composed of mostly harzburgite (forsteritic olivine $Fo_{93-95}$ and enstatite Mg# = 93 - 95) with an outer dunite portion, and a reacting basaltic zone. These zones are similar to those produced in the anhydrous peridotite-basalt systems at 0.1 MPa – 0.8 GPa (Fisk, 1986; Morgan & Liang, 2003). An addition of water at 1.0 GPa at the conditions of the basaltic melt saturation with water fluid phase likely decreases the timescale required for the serpentinite assimilation from 9h in the P7 run to 3h in the P12 run (see **Table 2**).

Additionally, the hybrid run P37 sample is represented by predominant basaltic glass of $L_{bas}$ (72.8 wt% in the sample) with assemblage of interstitial glass ($L_{int}$, 0.4 wt%), forsteritic olivine (17.0), residual orthopyroxene (6.8), and accessory clinopyroxene and chromiferous magnetite with pores of fluid (3.0) (**Table 2**). The mixed sample number SB1 obtained at 0.2 GPa is represented by polyhedral olivine phenocrystals in matrix. In the matrix, this sample contains assemblage of clinopyroxene microphenocrysts, rims of the olivine phenocrysts and interstitial felsic glasses. Oxide minerals are represented by chromite microphenocrysts. The sample SBter1 is represented by homogeneous basaltic glass formed by complete hybridation of the starting basaltic liquid with serpentinite, whereas the sample SBbis3 contains residual crystallized aggregate of olivine. It is worth noting that the current experiments with predominant proportion of basaltic melt (72 – 88 wt%) longer than 5 – 8 h at 0.2 – 1.0 GPa produce the total assimilation of the serpentinite zone by the basaltic melt (**Fig. 1**), resulting in homogeneous Mg-rich basaltic glasses.

**Summary on the Melt Composition**

The olivine-rich zones host glass pockets of 10 to 200 µm in size. The composition of the interstitial glasses produced in the shortest hybrid runs at 0.5 to 1.0 GPa varies from basaltic, basaltic andesite to andesitic (**Fig. 2**). The interstitial dacitic melts even richer in silica (up to 66 to 71 wt%) were produced at 0.2 GPa pressure compared to those obtained in the 0.5 – 1.0 GPa pressure (**Fig. 2**) and are discussed elsewhere (Borisova et al., 2020). In this work, we pay more attention to the mechanisms and to the rates of assimilation relevant to the magmas interacting with hydrated mantle lithosphere in oceanic setting. The interstitial melts produced at 0.2 to 1.0 GPa are formed close to equilibrium with olivines of the olivine-rich zones (**Fig. 3**). The major element composition of the basaltic melts produced due to the bulk serpentinite dissolution in the longest experiments (48h at 0.2 GPa, 8 h at 0.5 GPa and 3 h at 1.0 GPa pressure) indicates a strong contribution of the serpentinite on the basalt chemistry. Indeed, the final products of the kinetic series at 0.2 – 1.0 GPa are represented by homogeneous Mg-rich basaltic glasses (50 - 52 wt% $SiO_2$, 12 – 13 wt.% MgO) with ~500 ppm Cr and 140 – 200 ppm Ni (**Tables 3, 4, A1**). They contrast with the starting MORB (50 wt% $SiO_2$, MgO = 8.2 wt.%, Cr = 275 ppm, Ni = 129 ppm) due to the complete dissolution of the Mg, Cr and Ni-enriched serpentinite (**Figs. 2, 3**). The homogeneous basaltic glasses present lower $TiO_2$ contents (1.22 – 1.36 wt%) and similar primitive-normalized $(La/Sm)_n$ (1.1 – 1.5) and $(La/Yb)_n$ (1.6 – 2.0) ratios compared to those of the starting basaltic melt (1.45 wt%, 1.1 and 1.6, respectively) due to the low content of these elements in the serpentine (**Table 4**). Thus, the chemical impact of serpentinite on the final basaltic melt as a result of the bulk assimilation is a dilution in incompatible elements (e.g., Ti) contents and an enrichment in compatible Cr, Ni and Mg elements, as well as a slight enrichment in Si contents.

## Mechanism and Rate of the Hydrated Mantle Assimilation

### Mechanism of the Basaltic Melt-Hydrated Peridotite Reaction

Hybrid and mixed experiments on identical materials with high basaltic melt to serpentinite rock ratio (>2) performed at different run durations allowed determination of the reaction mechanism and the assimilation rate. The transformation of serpentinite to dehydrated harzburgite with pore fluid is demonstrated in the zero-time experiment at 1300°C and 0.5 GPa (**Fig. 1**, **Table 2**). The serpentinite in the shortest runs produces forsteritic olivine ($Fo_{91-95}$), enstatite (Mg# = 94 – 97) and chromite and/or chromiferous magnetite with Cr# = 7 - 89 and Mg# = 27 – 54 (**Fig. 1**, **Table 2**), similarly to the results of Chepurov et al. (2016). The phase composition diagram and MgO-$SiO_2$ plot (**Figs. 1, 3**) suggest that the basaltic melt-serpentinite reaction at 0.2 – 1.0 GPa is controlled by the following stages: (1) transformation of serpentinite to chromite-bearing harzburgite (crystallization of forsteritic olivine and enstatite with accessory chromite and liberation of pore fluid), (2) incongruent partial melting of the harzburgite and formation of olivine-rich zones with hydrous interstitial melts. Subsequently, the mechanism involves (3) formation of external chromite-bearing dunite zone due to incongruent dissolution of orthopyroxene associated with diffusive exchange between the reacting basaltic melt and the interstitial melts, (4) partial dissolution of the chromite-bearing dunite in the initial basaltic melt and finally (5) the total assimilation of the chromite-bearing dunite and formation of the high-Mg hydrous hybrid basaltic melt (**Tables 2, 3**). Moreover, the homogeneous hybrid basaltic melts produced in the longest experiments at 0.2 to 1.0 GPa are enriched in Mg and Cr due to the total assimilation of serpentinite (**Table 4**). Tholeiitic basaltic melts reacting with and assimilating the hydrous peridotite become eventually saturated in olivine, chromite and orthopyroxene. We suggest that the reaction of such depleted basaltic melt with a peridotite would produce chromite-bearing harzburgite through reactive porous flow. The mechanism observed in our work contrasts with reactive fractionation reported by Van den Bleeken et al. (2010; 2011) at 0.65 – 0.80 GPa in dry conditions. These authors observed plagioclase and orthopyroxene crystallization and formation of plagioclase-bearing peridotite due to reactive porous flow of dry tholeiitic basaltic melt through anhydrous peridotite at 1170 - 1320°C.

Fractional crystallization of the hybrid, depleted and chromite-saturated basaltic melts can produce chromitites. Indeed, according to the recently proposed model of the chromitite genesis, the bulk serpentinite assimilation by MORB basaltic melt is the main factor responsible for the massive chromite crystallization (Borisova et al., 2012a) at the oceanic Moho mantle-crust transition zone. Our experiments demonstrate that formation of the hybrid mid-ocean ridge basaltic melt saturated with chromite is possible at conditions of predominant proportion of basaltic melt (above 70 wt%) and, therefore, due to high melt/rock ratio (>2) at 0.2 to 1.0 GPa pressure. The main physico-chemical parameters controlling the chromite crystallization are the presence of aqueous fluid or/and hydrous basaltic melts into the reactive system (Borisova et al., 2012a; Johan et al., 2017; Zagrtdenov et al., 2018). The fluid presence at 0.2 GPa pressure is also necessary condition for the chromite concentrating by a physical, not a chemical process due to surface tension of the fluid, which is sufficient to maintain dispersed chromite crystals inside the fluid upon the chromite crystallization (Matveev & Balhaus, 2002). Additionally, an initial stage of the basaltic melt reaction with dehydrated serpentinite may result in formation of chromite-bearing harzburgite and dunite. The first direct evidence that the initial stage really happens at the 0.2 GPa mantle-crust transition zone is occurrence of chromite-hosted silica-rich inclusions of the Oman ophiolite chromitite ore bodies (Rospabé et al., 2019b).

## Assimilation Rate of the Hydrated Mantle Lithosphere

The calculated average rate of the serpentinite assimilation by the basaltic melt in the experiments without additional water is $4.3 \times 10^{-10}$ $m^2/s$ (**Table 5**). It is surprisingly similar to the assimilation rate by basaltic melt in the runs with an additional water at conditions of the melt saturation with an aqueous fluid (~$4.0 \times 10^{-10}$ $m^2/s$). **Figure 4** demonstrates that the serpentinite assimilation rate by dry basaltic melt measured at 0.5 GPa is progressively decreasing during the experimental run. This may be explained by approaching equilibrium, in accordance with principles of the chemical kinetics. The calculated rates are at least one order of magnitude higher compared to $10^{-12}$ - $10^{-11}$ $m^2/s$ for the basalt interacting with anhydrous harzburgite at 0.6 GPa (Morgan and Liang, 2003). The assimilation rate estimated in our work is comparable to the silica diffusivity in a hydrous basaltic melt (Zhang et al. 2010). Similarly, the reaction rate established by Morgan and Liang (2003) in anhydrous system is mostly comparable to the silica diffusivity in dry basaltic melt. Since both reaction rates are controlled by the silica diffusion, the difference in the reaction rates is related to the well-known promoting effect of $H_2O$ on the silica diffusion in silicate melts (e.g., Zhang et al. 2010 and references therein). Thus, the serpentinite assimilation by basaltic magma may proceed at least 10 times faster than the formation of dunitic reaction margins (i.e. “dyke walls”) in the oceanic lithosphere during transport of dry basaltic melt (Morgan and Liang, 2003) at 0.6 – 0.8 GPa. Additionally, the newly produced hybrid basaltic melts which become highly saturated in olivine, chromite and, likely, in orthopyroxene would produce plagioclase-free chromite-bearing harzburgite through reactive porous flow in a peridotite. This process may be expressed in nature by formation of chromite-bearing harzburgitic rather than dunitic channels of depleted hybrid magmas enriched in Si, Mg, Cr, and $H_2O$.

**Comparison to Chemistry of Oceanic Magmas and Glasses**

**Figure 2** illustrates a huge range of the major elements, water and Cr contents in natural MORB, depleted basaltic and oceanic boninite glasses at wide range of $SiO_2$ (40 – 69 wt.%) contents taken from the PetDB database (Lehnert et al., 2000). These compositional variations in the oceanic melts are generally attributed to varying partial melting and fractional crystallization of basaltic and boninite magmas as well as to the reactive basaltic melt percolation through anhydrous peridotite (e.g., Van den Bleeken et al., 2010; 2011). However, the compositional similarity of the experimental glasses produced due to serpentinite assimilation by basaltic melt to the compositional field of the natural oceanic glasses indicates that the composition of the oceanic magmas may be also controlled by the melt reaction with hydrated peridotite during the magma percolation to the seafloor. The elevated water contents in the experimental liquids related to the serpentinite dehydration at magmatic conditions is a main difference between the experimental liquids and natural glasses. This difference may be explained by the natural basaltic melt degassing upon its residence and transport to the surface. Indeed, the highest known water contents of the Kane Megamullion south of the Kane Fracture Zone along the Mid-Atlantic Ridge were recorded in MORB samples (up to 1.8 wt% in glass, and up to 2.7 wt% in bulk rocks) (Ciazela et al., 2017). These MORB magmas are considered to assimilate significant volume of the host serpentinite at lithospheric conditions, whereas most MORB glasses sampled at the surface are degassed.

It is now established that hydrothermal circulation during oceanic spreading reaches mantle peridotite at and below the petrologic Moho (Python et al., 2007; Rospabé et al., 2017; 2019a,b). The serpentinized peridotite or/and dehydrated serpentinite mantle are characterized by an excess of $H_2O$, Cl (e.g., Bonifacie et al., 2008), $^{4}$He, $^{36}$Ar (Kendrick et al., 2013) and radiogenic $^{87}$Sr/$^{86}$Sr (e.g., Harvey et al., 2014). The oceanic melts affected by assimilation of serpentinized mantle thus may be recognized by an excess of Si, Cr, Mg, $H_2O$, Cl, and radiogenic $^{87}$Sr/$^{86}$Sr ratio compared to those of typical MORB. It should be noted that the highest measured Cr contents in the chromite-hosted melt inclusions of Husen et al. (2016) are overestimated due to secondary fluorescence effects of nearby chromite (Borisova et al. 2018), whereas these effects are absent during EPMA of natural glasses without any trace of chrome-rich minerals. The hybrid oceanic magmas thus may be distinguished from the original MORB melts by non-mantle isotopic compositions of O, H, Cl, He, Ne, Ar. For example, elevated Cl contents and atmospheric Ne and Ar in the Mid-Atlantic Ridge basaltic glasses (Stroncik & Niedermann, 2016) are explained by the shallow-level assimilation of a seawater-sourced component. Low $^{3}$He/$^{4}$He ratios of the Southwest Indian Ridge MORB interpreted by Georgen et al. (2003) as recycled lithospheric material in the upper mantle source region, may be ascribed to the assimilation of the serpentinized lithospheric mantle by the oceanic tholeiitic basalt. Our data confirm the hypothesis of Benoit et al. (1999) and Nonnotte et al. (2005), based on $^{87}Sr/^{86}Sr$-$^{143}Nd/^{144}Nd$ ratios, that petrogenesis of orthopyroxene-rich primitive and depleted oceanic cumulates from the Mid Atlantic Ridge (DSDP Site 334) and at the periphery of the Maqsad paleo-diapir in the Oman ophiolite may be due to hybridization (or mixing) between tholeiitic basaltic melts and liquids issued from melting of the serpentinized lithospheric mantle. Similarly, the radiogenic $^{87}Sr/^{86}Sr$ coupled with an excess of $H_2O$, $SiO_2$, halogens, $^{4}He$, $^{36}Ar$ in oceanic basalts, related cumulates and associated mantle peridotites is undisputable evidence for their origin due to serpentinization of the mantle lithosphere (Neumann et al., 2015) and the basaltic melt reactions with the serpentinized mantle rather than due to partial melting of deep-seated serpentinized mantle (DMM, HIMU, e.g., Kendrick et al., 2017).

The effect of the altered lithosphere assimilation is likely widespread given the elevated chlorine contents (> 50 ppm) and the resulting high Cl/K ratios (> 0.09) in the MORB glasses well above Cl of 50 ppm content and global mantle Cl/K of 0.09, respectively (Stroncik & Nidermann, 2016; van der Zwan et al., 2017). Clinopyroxene-melt thermobarometry yields lower crustal-upper mantle crystallization/assimilation depths of 10 – 13 km. The Cl-excess is observed in both slow-spreading and fast-spreading ridges (van der Zwan et al., 2017 and references therein). The mechanism of such Cl contamination process may be either assimilation of the country rocks containing pore brine fluid, or brine in fluid inclusions or partial melting of the altered lithosphere, summarized by van der Zwan et al. (2017) to the 'assimilation of hydrothermally altered lithosphere'. Recent investigation of MOR gabbro hosted by the Kane Megamullion serpentinized mantle supports the existence of the basaltic melt reaction with the serpentinizeed mantle enriching the final melt in water (up to 2.7 wt% in bulk rocks), triggering sulfide saturation, clinopyroxene crystallization before plagioclase and lowering solidus temperature down to 840°C. The effect of the melt-rock interaction allows even larger amount of melt to react with the channels or dyke walls (Ciazela et al., 2017; 2018). These facts and the well-documented contact metamorphic rock consisting of antigorite, tremolite, chlorite and prehnite suggest that the mafic melt-serpentinite reaction in fact takes place (Ciazela et al., 2017; 2018), especially in slow-spreading ridges (van der Zwan et al., 2017).

Additionally, ultra-depleted MORB and related variously depleted melts are produced beneath mid-ocean ridges, likely at 0.5 – 2.0 GPa pressure in assemblage with forsteritic olivine ($Fo_{82-91}$ mol.%) according to Ross and Elthon (1993), Sobolev and Shimizu (1993) and Hunsen et al. (2016). Routinely, origin of such ultra-depleted and differently depleted MORB melts has been attributed to an effect of critical (continuous) melting with formation of lherzolite/harzburgite residue. Chemical similarity of our experimental Cr-Mg-$H_2O$-rich melts and associated high-Mg silicates such as forsteritic olivine ($Fo_{91-95}$ mol.%) to those associated to forsteritic olivine ($Fo_{82-91}$ mol.%) (e.g., Sobolev and Shimizu, 1993; Husen et al., 2016) implies that origin of some part of these variously depleted oceanic basaltic melts may be attributed to the reaction between tholeiitic basaltic melt and the serpentinized mantle rather than to fractional “dynamic” melting of mantle peridotite.

## Integrated Model of Basaltic Melt - Hydrated Peridotite Reaction

It is likely that the serpentinized lithosphere assimilation may be stronger along slow- and ultraslow-spreading ridges due to faults rooting deeper, providing pathway for hydrothermal fluids (e.g., Mével & Cannat, 1991; van der Zwan et al., 2017 and references therein), although there are clear evidence for elevated concentrations of Cl in the MORB magmas of fast-spreading ridges (e.g., France et al., 2009) and in small oceanic island magmas (e.g., Dixon et al., 2008 and references therein). Several examples suggest that the assimilation is favorable in the slow-spreading settings where the source of the Cl contamination lies deep in the magmatic system, although Cl excess in magmas and evidence for the assimilation of the hydrothermally altered lithosphere are observed in all types of ridges, whatever their spreading rate (Ciazela et al., 2017; van der Zwan et al., 2017).

Our experimental work on basaltic melt-serpentinite interaction provides convincing evidence that several types of the oceanic lavas and cumulates can be produced by reaction of a typical mid-ocean ridge basaltic melt with serpentinized lithospheric mantle. Hybrid experiments suggest multi-stage reactions with serpentinized mantle (**Figure 5)**: (i) an initial stage of primitive basalt percolation into the serpentinized mantle; (ii) serpentinite dehydration and transformation to chromite-bearing harzburgite containing pore fluid and generation of hydrous interstitial melts variously enriched in silica; and (iii) the final production of hybrid melts depleted in incompatible elements in association with chromite-bearing harzburgite or dunite due to progressive incongruent dissolution of orthopyroxene and reaction with basaltic melt at 0.2 – 1.0 GPa pressure. Since the observed major and trace element composition of the produced melts depends on the reaction pressure and duration (e.g., **Figs. 1 - 3**), our data infer that chemical evolution of the oceanic basaltic magmas depends on (1) the depth of their interaction with the overlying oceanic lithospheric mantle serpentinized by seawater-derived fluids and (2) the rate of the basaltic melt transport from their upper mantle source, i.e., how long the oceanic melts interacted with the serpentinized lithospheric mantle.

## Conclusions

1) The hybrid and mixed experiments performed at 0.2 to 1.0 GPa pressures on interaction between basaltic melt and serpentinite provide convincing evidence that generation of depleted MORB melts, high-Mg-Cr cumulates, chromitites, and oceanic boninites and andesites can be reliably explained by the efficient reaction of initially anhydrous basaltic melts with the serpentinized lithospheric mantle. Our data infer that chemical evolution of the oceanic magmas depends on which depth and how long the oceanic basaltic melts interacted with the hydrated mantle lithosphere. Our work determines physical-chemical conditions at which reaction of the hydrated peridotite (or serpentinite) with basaltic liquid can lead to massive chromite crystallization at 0.2 GPa pressure according to model of Borisova et al. (2012a).

2) The effect of the reaction of tholeiitic basaltic melt with hydrous peridotite contrasts strongly with that observed on anhydrous peridotite at 0.65 – 0.80 GPa by Van den Bleeken et al. (2010; 2011). The main difference is the total assimilation of hydrous harzburgite by the basaltic magma and a complete absence of plagioclase in the reacted harzburgite. That is likely due to the effect of elevated water contents into the hybrid systems suppressing crystallization of plagioclase. Tholeiitic basaltic melts reacting with and assimilating the hydrous peridotite become more saturated in olivine, chromite and orthopyroxene and may produce chromite-bearing harzburgite upon reactive melt transport in the upper mantle.

3) The rate of the bulk serpentinized peridotite assimilation by tholeiitic basaltic melt (4.0 – $4.3 \times 10^{-10}$ $m^2/s$) is controlled by the silica diffusion in hydrous basaltic melts. The calculated rates of the serpentinite assimilation are at least one order of magnitude higher than $10^{-12}$ - $10^{-11}$ $m/s^2$ for dry basaltic melt interacting with anhydrous harzburgite and producing dunite channels due to reactive porous flow of the basaltic melt (Morgan and Liang, 2003). The bulk assimilation of serpentinite by basaltic melt may happen at conditions of high melt/rock ratio (>2) and the predominant proportion of basaltic melt above 70 wt% in the hybrid system.

4) Our study challenges the routine interpretation of variations in chemical and isotopic composition of oceanic lavas (e.g., MORB and OIB) in terms of deep mantle plume source heterogeneities or/and mechanism of partial melting.

## Acknowledgements

We thank Editor Prof. Yigang Xu, reviewer Prof. Tomo Morishita and one anonymous reviewer for comments and suggestions, which helped to improve the manuscript. This work was supported by the Institut Carnot ISIFoR, Deutsche Forschungsgemeinschaft (DFG) German Research Foundation, the University of Bayreuth, and AST Planets of the Observatoire Midi-Pyrénées. We acknowledge the access to the experimental facilities of the Bavarian Research Institute of Experimental Geochemistry and Geophysics (BGI) and to the European Synchrotron Radiation Facility (ESRF). We thank C. McCammon, H. Keppler and N. Dubrovinskaia for advice and assistance with the experiments, M. Munoz for help with the XANES data analyses, Z. Morgan, Y. Liang and M. Rabinowicz for discussions on the basalt-peridotite reactions that helped improve this article.

## Author Contributions

A.Y. Borisova, M.J.T., G.C. and O.G.S. developed the conceptual idea of the study; A.Y. Borisova and N.R.Z., S.S. prepared and conducted high T-P runs at BGI; N.R.Z., O.G.S., A.Y. Bychkov prepared and performed experiments at IEM; G.S.P. and N.R.Z. carried out XAS measurements; K.P.J., B.S. and U.W. performed LA-ICP-MS analyses. A.Y. Borisova and N.R.Z. performed microanalytical measurements and mapping using EPMA and A.Y. Borisova, and G.C. developed geological applications; all authors contributed to data interpretation and manuscript writing.

# FIGURES

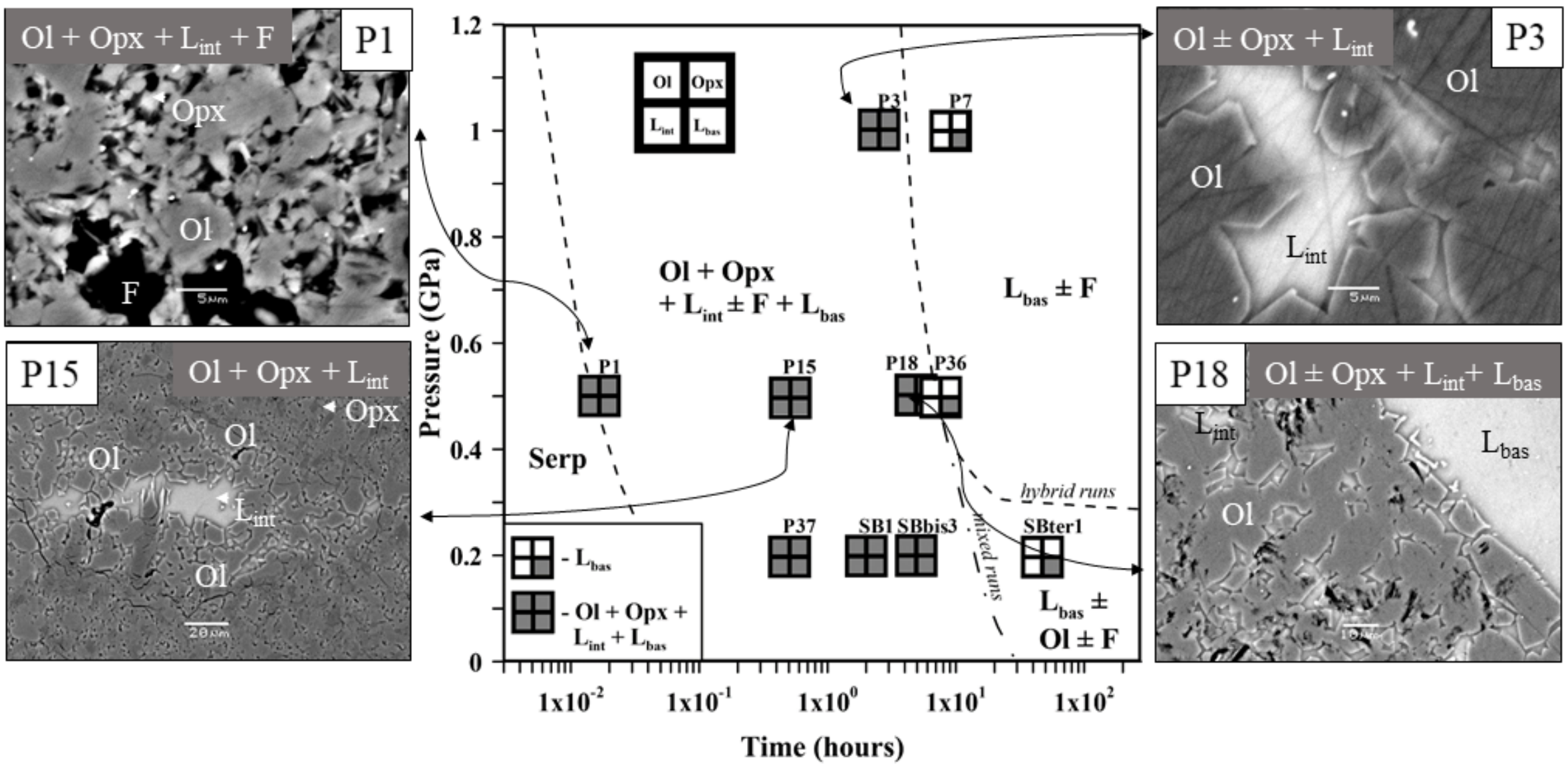

**Figure 1.** Pressure (in GPa) versus experimental run duration (in hours) diagram in the experimental basaltic melt-serpentinite system (without additional water) showing mineral and glass assemblages produced in hybrid and mixed runs at 1250 – 1300 °C. The diagram and the back-scattered electron images of the samples show the main melt and mineral phases observed after quenching: $L_{bas}$ – zone of hydrous basaltic glass; $L_{int}$ – interstitial glass; Opx – orthopyroxene; Ol – forsteritic olivine; F – fluid; Serp – serpentine, mostly antigorite minerals issued from the starting serpentinite. The run numbers and corresponding run products are given in Table 2.

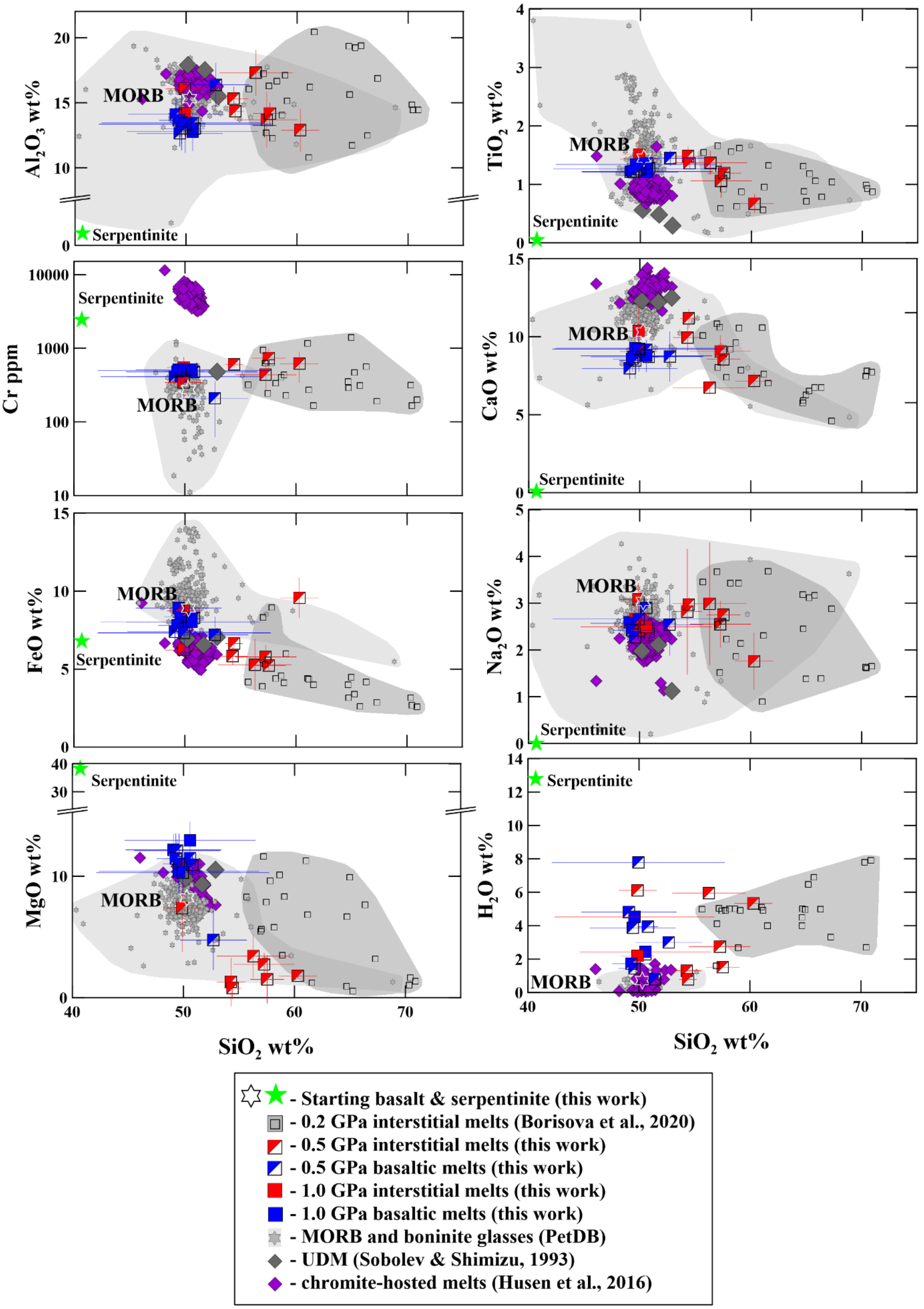

**Figure 2.** $Al_2O_3$, $TiO_2$, FeO, MgO, CaO and $Na_2O$ (in wt%), Cr (ppm) and $H_2O$ (in wt%) contents versus $SiO_2$ contents (in wt%) in the experimental glasses. Composition of the experimental glasses produced from 0.2 to 1.0 GPa is compared to the starting basalt (MORB)

and the serpentinite compositions as well as to the compositional fields of natural tholeiitic and oceanic boninitic glasses (from database of PetDB, Lehnert et al., 2000), ultra-depleted melts marked as UDM (Sobolev and Shimizu, 1993) and chromite-hosted melts (Husen et al., 2016) which are differently depleted MORB melts. The recalculated melts obtained experimentally at 0.2 GPa pressure are after Borisova et al. (2020). $H_2O$ contents in the experimental glasses are values calculated from the EPMA. All experimental data are available in Tables 3 and A1.

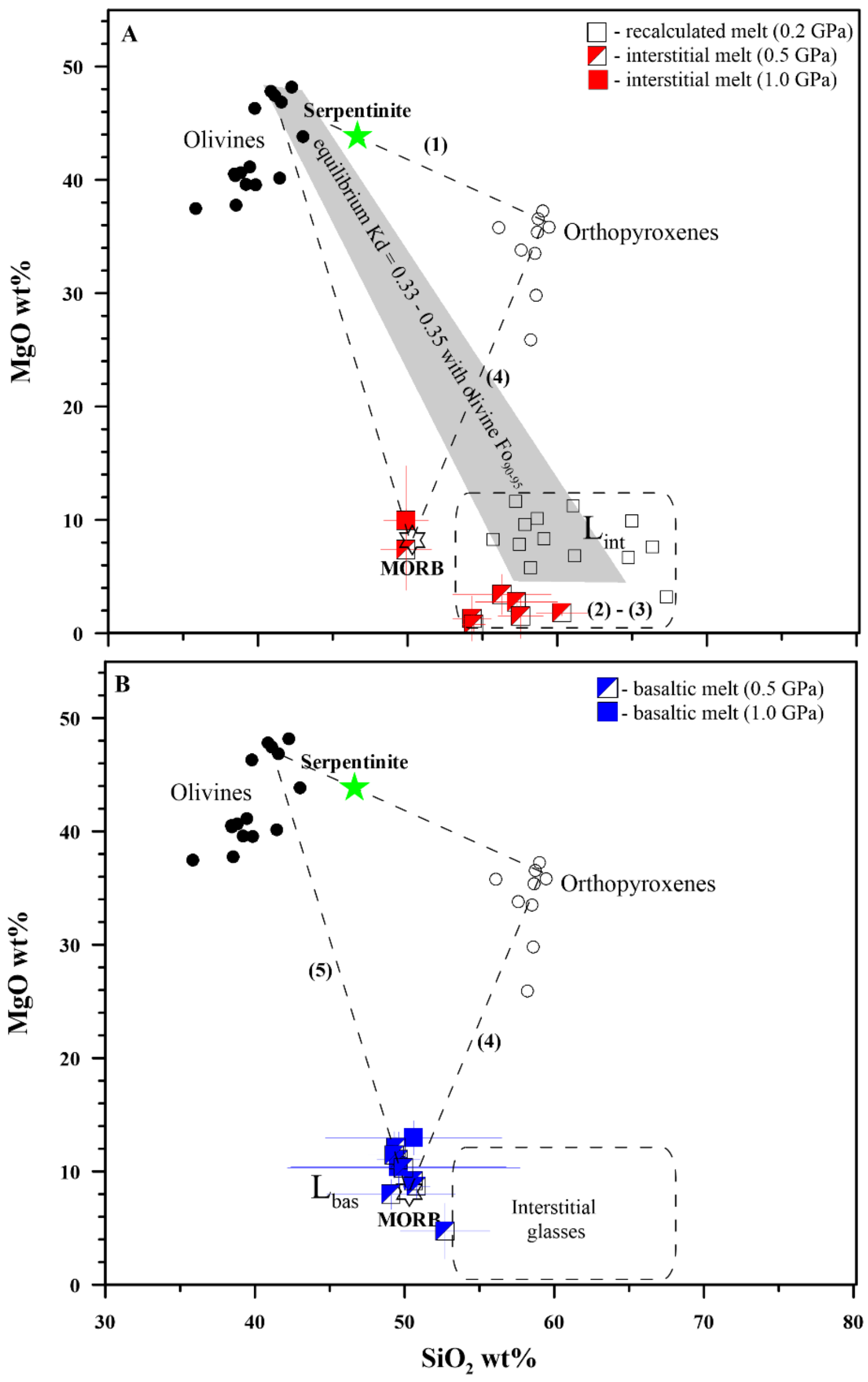

**Figure 3 (a,b).** Mechanism of the basaltic melt interaction with serpentinite. Five steps of the interaction have been distinguished: (1) dehydration and transformation of serpentinite to Cr-rich spinel-bearing harzburgite (crystallization of olivine and orthopyroxene) with appearance of pore fluid, (2) incongruent melting of the harzburgite and formation of associated hydrous interstitial melts in harzburgite/dunite, (3) progressive formation of external dunite zone due to orthopyroxene dissolution and diffusive exchange between the external basaltic melt and interstitial melts and (4) dissolution of the dunite in the basaltic melt and finally (5) bulk dissolution of the dunite and formation of hybrid basaltic melts. $L_{int}$ and $L_{bas}$ are interstitial and basaltic melts, respectively (see Table 2). Equilibrium Kd are calculated as theoretical FeO-MgO partition coefficient between olivine ($Fo_{90-95}$ mol%) and co-existing interstitial melt of 57 to 62 wt% $SiO_2$ at 0.5 GPa pressure according to Toplis (2005). The recalculated melts obtained experimentally at 0.2 GPa pressure are after Borisova et al. (2020). All experimental data are available in Tables 3 and A1.

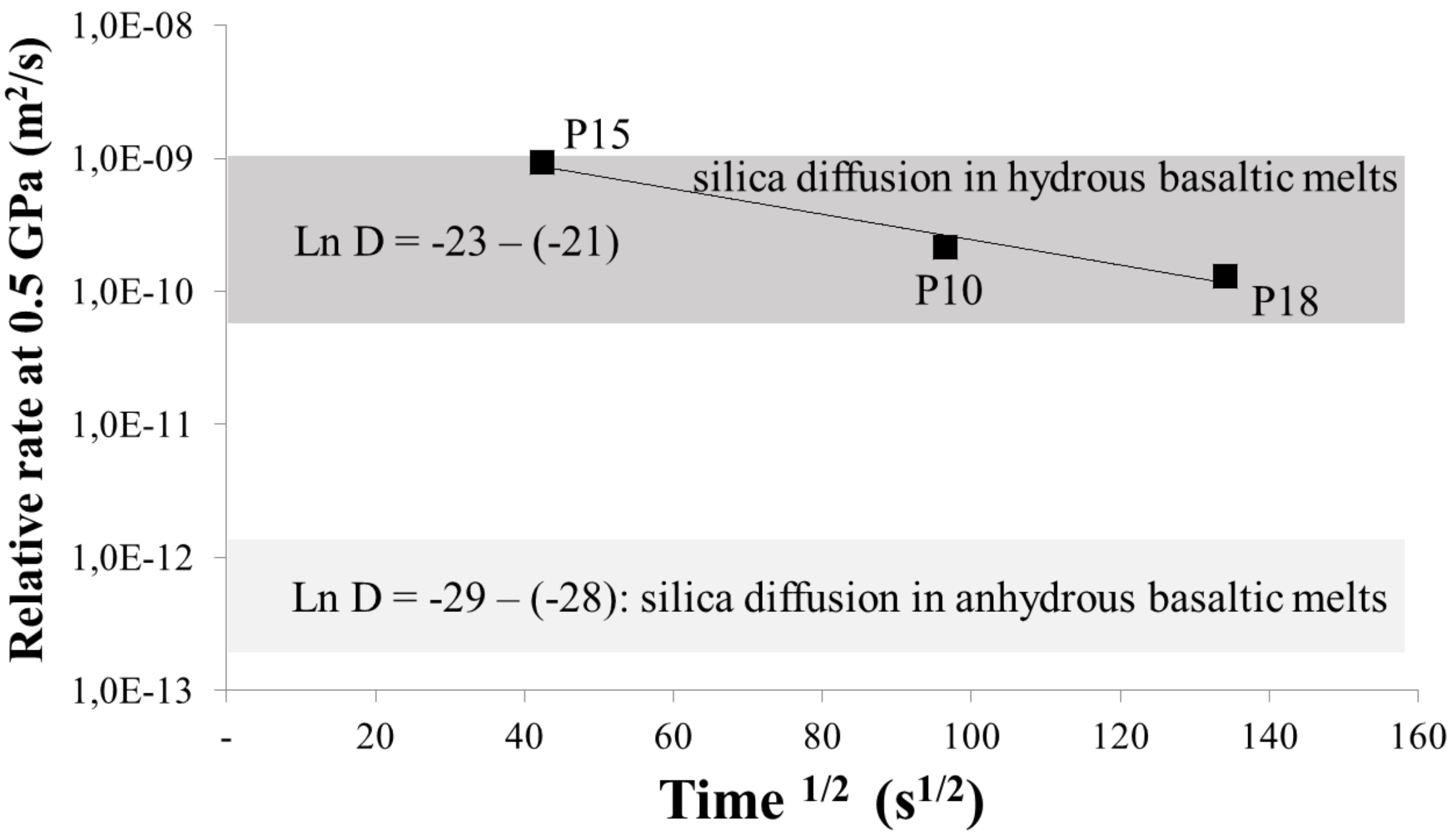

**Figure 4.** 'Relative' rates (in $m^2/s$) of the serpentinite assimilation by the basaltic melt at 1300°C and 0.5 GPa versus the run duration (in $s^{1/2}$). Table 5 demonstrates the applied calculation method for the "relative dissolution rate".

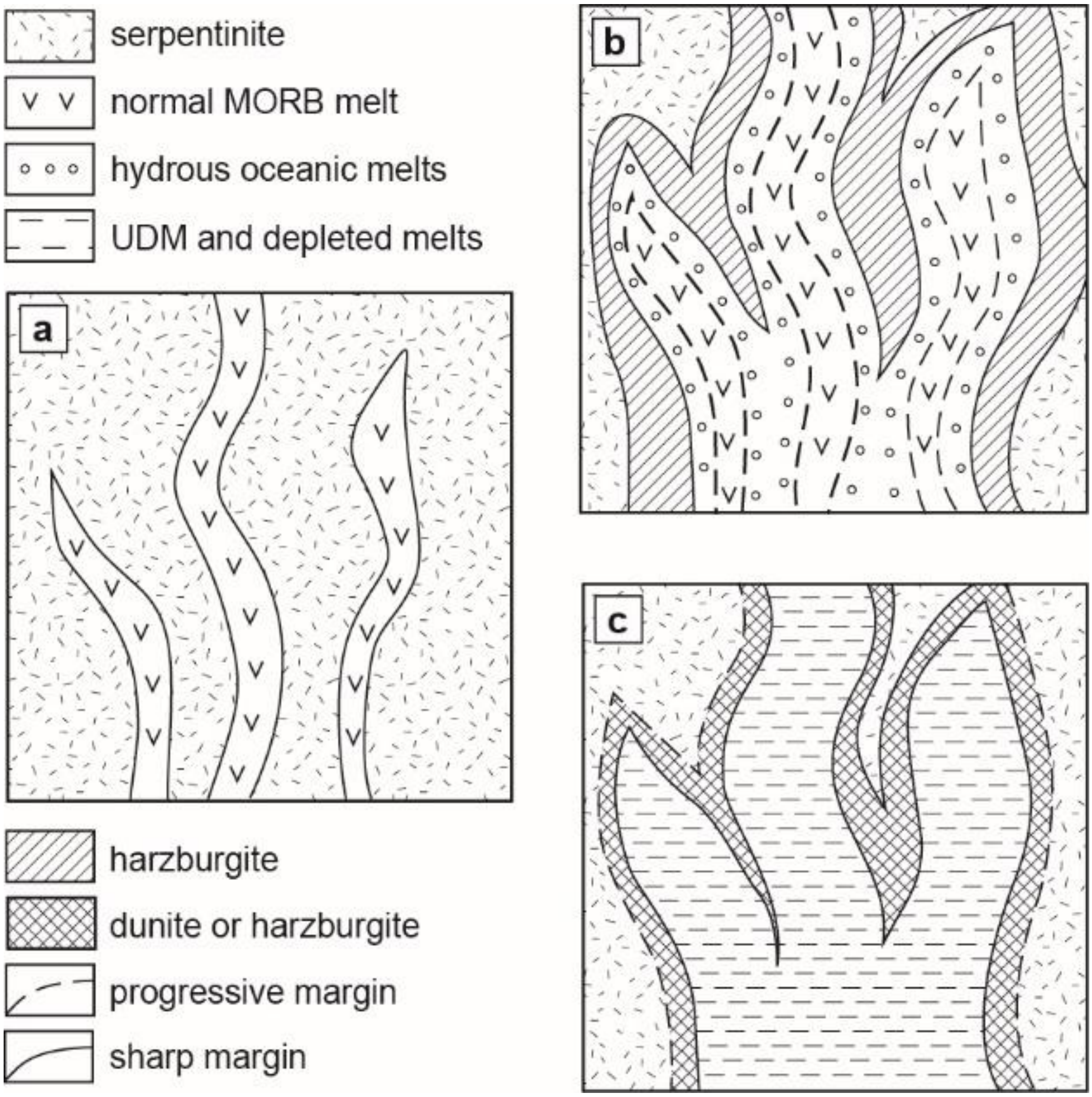

**Figure 5.** Model of the reaction between "normal" oceanic basaltic melt and the hydrated (serpentinized) mantle lithosphere. (a) The initial stage of the mafic melt percolation into the hydrated mantle and formation of contact metamorphic harzburgite containing pore fluid; (b) formation of harzburgite wall rocks in assemblage with interstitial hydrous silica-enriched (basaltic to dacitic) melts and diffusive homogenization between the initial oceanic basaltic (e.g., MORB, OIB) melt and the produced hydrous silica-enriched melts. At this stage, formation of peridotite xenoliths containing silica-rich inclusions is highly probable. (c) The final production of variously depleted hybrid melts locally associated with the chromite-bearing dunite/harzburgite. The bulk assimilation of serpentinite by basaltic melt may happen at conditions of high melt/rock ratio and the predominant proportion of basaltic melt above 70 wt% into the hybrid system.

**Table 1.** Starting material composition

| Sample | Basalt* | TSL-19 |
|---|---|---|
| $SiO_2$, wt% | 50.31 | 40.69 |
| $TiO_2$ | 1.45 | D.L. |
| $Al_2O_3$ | 15.31 | 0.88 |
| $Fe_2O_{3total}$ | 9.9 | 7.53 |
| MnO | 0.17 | 0.11 |
| MgO | 8.21 | 38.24 |
| CaO | 10.29 | 0.06 |
| $Na_2O$ | 3.04 | D.L. |
| $K_2O$ | 0.31 | D.L. |
| $P_2O_5$ | 0.18 | 0.05 |
| LOI | -0.22 | 11.55 |
| Total | 98.93 | 99.1 |
| $H_2O$ total, wt% | 0.71 | 11.91 |
| Ba, ppm | 75.8 | D.L. |
| Ce | 16.8 | D.L. |
| Cr | 275 | 2417 |
| Dy | 5 | 0.049 |
| Er | 3.04 | 0.042 |
| Eu | 1.33 | 0.008 |
| Gd | 4.34 | 0.023 |
| Hf | 2.65 | D.L. |
| Ho | 1.12 | 0.012 |
| La | 6.44 | D.L. |
| Lu | 0.445 | 0.009 |
| Nb | 7.23 | D.L. |
| Nd | 12.4 | D.L. |
| Ni | 129 | 1901 |
| Pr | 2.51 | D.L. |
| Rb | 7.19 | D.L. |
| Sm | 3.75 | D.L. |
| Sr | 148 | 2.349 |
| Tb | 0.75 | 0.005 |
| Th | 0.63 | D.L. |
| Tm | 0.436 | 0.007 |
| U | 0.79 | 0.627 |
| Y | 29.1 | 0.266 |
| Yb | 2.91 | 0.05 |
| Zr | 94.3 | D.L. |

*Basalt – mid-ocean ridge basaltic glass (Sushchevskaya et al. 2000) used in the experiments; TSL-19 - serpentinite (Deschamps et al., 2010) used in the experiments. D.L. – concentrations are below detection limit.

**Table 2. Conditions and products of the experimental interaction between basaltic melt and hydrated peridotite at 0.2 – 1.0 GPa**

| N° | Run | Pressure (GPa) | Temperature (°C) | Capsule material | Time (h) | Percentage of the starting components (wt%) | | | $H_2O$* (wt%) ($\alpha_{H2O}$) | Run products with phase proportions (wt%) | Resulting oxygen fugacity QFM (X $Fe^{3+}$)XANES $$ | Kd |
|---|---|---|---|---|---|---|---|---|---|---|---|---|
| | | | | | | Serpentinite[a] | Basalt$ | Water$ | | | | |
| 1 | P37 | 0.2 | 1250 | $Au_{80}Pd_{20}$ | 0.5 | 28.2 | 71.8 | - | - | L bas (72.8) + L int (0.4) + Ol (17.0) + Opx (6.8) + Cpx + ChrMgt (0.0) + F (3.0) | - | - |
| 2 | SB1$$$ | 0.2 | 1250 | $Au_{80}Pd_{20}$ | 2 | 19.95 | 80.05 | - | - | Ol (39.6) + Cpx (22.3) + Chr (0.8) + $L_{int}$ (34.0) + F (3.3) | - | - |
| 3 | SBbis3$$$ | 0.2 | 1250 | $Au_{80}Pd_{20}$ | 5 | 19.95 | 80.05 | - | - | Ol + $L_{bas}$ + F | - | - |
| 4 | SBter1$$$ | 0.2 | 1250 | $Au_{80}Pd_{20}$ | 48 | 19.95 | 80.05 | - | - | $L_{bas}$ + F | - | - |
| 5 | P1 | 0.5 | 1300 | $Au_{80}Pd_{20}$ | 0.02 | 19.2 | 80.8 | - | 2.8 | $L_{int}$ + Ol + Serp+ Opx + Cpx + Chr + $L_{bas}$ | 0.14 - 0.19 QFM (2.9 – 3.6) | 0.36 ± 0.05 |
| 6 | P15 | 0.5 | 1300 | $Au_{80}Pd_{20}$ | 0.5 | 15.7 | 84.3 | - | 2.2 | $L_{int}$ (3) + Ol (13) + Opx (3) + ChrMgt + $L_{bas}$ (81) | - | - |
| 7 | P10 | 0.5 | 1300 | Pt | 2.6 | 12.9 | 87.1 | - | 1.8 | $L_{int}$ + Ol + Cpx + Chr + $L_{bas}$ | QFM (1.5 – 3.0) | - |
| 8 | P18 | 0.5 | 1300 | $Au_{80}Pd_{20}$ | 5.0 | 13.4 | 86.6 | - | 1.8 | $L_{int}$ (2) + Ol (11) + Opx (2) + Chr + $L_{bas}$ (85) | - | - |
| 9 | P36 | 0.5 | 1250 | $Au_{80}Pd_{20}$ | 8.0 | 14.6 | 85.4 | - | 2.0 | $L_{bas}$ | | - |
| 10 | P20 | 0.5 | 1300 | $Au_{80}Pd_{20}$ | 0.5 | 14.6 | 77.6 | 7.8 | (1)* | $L_{int}$ (2) + Ol (11) + Opx (3) + Chr + $L_{bas}$ (84) + F | - | - |
| 11 | P21 | 0.5 | 1300 | $Au_{80}Pd_{20}$ | 2.5 | 11.8 | 78.7 | 9.5 | (1) | $L_{int}$ (2) + Ol (11) + Opx (3) + Chr + $L_{bas}$ (85) + F | - | - |
| 12 | P26 | 0.5 | 1300 | $Au_{80}Pd_{20}$ | 5.0 | 13 | 80.6 | 6.4 | (1) | $L_{int}$ (0.3) + Ol (4) + Opx (0.8) + Chr + $L_{bas}$ (94) + F | - | - |
| 13 | P3 | 1.0 | 1300 | $Au_{80}Pd_{20}$ | 2.5 | 15.3 | 84.7 | - | 2.1 | $L_{int}$ (3) + Ol (11) + Opx (4) + ChrMgt + $L_{bas}$ (83) | - 0.13 QFM (2.5) | 0.33 ± 0.05 - |
| 14 | P7 | 1.0 | 1300 | $Au_{80}Pd_{20}$ | 9.0 | 15.9 | 84.1 | - | 2.2 | $L_{bas}$ | - | - |
| 15 | P12 | 1.0 | 1300 | $Au_{80}Pd_{20}$ | 3.0 | 13.4 | 79.3 | 7.3 | (1) | $L_{bas}$ + F | 0.21 | - |

[a] Weight percent of the serpentine in the system is calculated as mass of serpentine divided by total mass of the all components: ($M_{Serp}$ / ($M_{Serp}$ + $M_{MORB}$) for P1, P15, P10, P18, P25, P3, P7 or $M_{Serp}$ / ($M_{Serp}$ + $M_{MORB}$ + $M_{H2O}$) for P20, P21, P26, P12, where $M_{Serp}$, $M_{MORB}$ and $M_{H2O}$ are mass of serpentine, basaltic glass and additional water, respectively.
[b] "$L_{int}$" = interstitial glass; "$L_{bas}$" = hydrous basaltic glass; "Serp" = serpentine; "Ol" = olivine; "Opx" = orthopyroxene; "Cpx" = clinopyroxene; "Amph" = amphibole; "Chr" = chromite; "ChrMgt" = chromiferous magnetite; "F" = water bubble(s).
[$] "Basalt" = Mid Atlantic Ridge basaltic glass; "Water" - additional water added to the starting system.
[$$] QFM = oxygen fugacity expressed in log units compared to the quartz-fayalite magnetite (QFM) mineral redox buffer according to Ballhaus et al. (1991).
[$$$] mixed runs performed at 0.2 GPa.
$(X\ Fe^{3+})^{XANES}$ – molar fraction of $Fe^{3+}$ compared to the bulk Fe in the sample measured by XANES.
Kd are theoretical FeO-MgO partition coefficient between olivine and co-existing melt (Toplis, 2005) compared to the experimental FeO-MgO partition coefficient between olivine and co-existing melt (0.23 ± 0.10) taking into consideration of $Fe^{II}$ and $Fe^{III}$ in the melt with assumed [X $Fe^{III}$] = 0.19 in the melt.; $H_2O$ wt% = 2 is suggested for the 0.5 - 1.0 GPa runs.

[*] Calculated maximal $H_2O$ content in the basaltic melt due to the water liberation from the reacting serpentinite. Water activity equal to 1 suggests the basaltic melt saturation with water fluid and the presence of the water fluid at the run conditions of 0.5 – 1.0 GPa pressure. The solid phase proportions were calculated based on the mass balance consideration.

## Table 3. Composition of mineral and glass phases from the 0.5 - 1.0 GPa basaltic melt-serpentinite reaction experiments

| Exp [a] | Phase [b] | $SiO_2$ (wt%) | $TiO_2$ (wt%) | $Al_2O_3$ (wt%) | $Cr_2O_3$ (wt%) | $FeO_{tot}$ [c] (wt%) | MnO (wt%) | MgO (wt%) | CaO (wt%) | NiO (wt%) | $Na_2O$ (wt%) | $K_2O$ (wt%) | $P_2O_5$ (wt%) | Cl (wt%) | Total (wt%) | $H_2O$ (wt%) | Cr (ppm) | Ni (ppm) | Mg# [d] | Cr# [e] |
|---|---|---|---|---|---|---|---|---|---|---|---|---|---|---|---|---|---|---|---|---|
| **P1** | L bas (36) | 52.8 ±3.1 | 1.46 ±0.14 | 16.39 ±1.69 | 0.03 ±0.02 | 7.3 ±1.17 | 0.13 ±0.04 | 4.75 ±2.46 | 8.74 ±1.61 | D.L. | 2.55 ±0.67 | 0.36 ±0.07 | 0.18 ±0.03 | 0.7 ±0.49 | 95.35 | 4.7 | 209 ±147 | D.L. | 50.84 ±9.09 | - |
| | L int (13) | 56.3 ±3.3 | 1.37 ±0.2 | 17.3 ±1.76 | 0.02 ±0.01 | 5.28 ±1.65 | 0.12 ±0.05 | 3.42 ±1.79 | 6.73 ±0.4 | D.L. | 2.99 ±1.31 | 0.36 ±0.1 | 0.18 ±0.04 | - | 94.04 | 6.0 | 112 ±93 | D.L. | 51.6 ±12.46 | - |
| | Ol (14) | 42.7 ±1 | D.L. | 0.45 ±0.44 | 0.2 ±0.23 | 4.34 ±0.91 | 0.12 ±0.03 | 50.82 ±1.91 | 0.29 ±0.25 | 0.35 ±0.02 | D.L. | D.L. | D.L. | - | 99.24 | - | 1358 ±1550 | 2721 ±191 | 95.41 ±1.02 | - |
| | Opx (7) | 58.9 ±0.8 | D.L. | 1.33 ±0.88 | 0.22 ±0.05 | 3.56 ±0.37 | 0.17 ±0.03 | 33 ±2.37 | 2.24 ±1.43 | 0.15 ±0.02 | 0.38 ±0.19 | 0.1 ±0.08 | D.L. | - | 100.0 | - | 1488 ±314 | 1181 ±165 | 94.46 ±1.13 | - |
| | Chr (1) | D.L. | 0.17 | 1.86 | 21.57 | 42.68 | 0.14 | 14.4 | D.L. | 0.36 | D.L. | D.L. | D.L. | - | 94.31 | - | 147556 | 2790 | 50.01 | 88.61 |
| **P15** | L bas (163) | 50.8 ±0.9 | 1.28 ±0.09 | 13.01 ±0.74 | 0.07 ±0.01 | 8.27 ±0.23 | 0.16 ±0.04 | 10.92 ±1.04 | 8.69 ±0.39 | D.L. | 2.39 ±0.12 | 0.27 ±0.05 | 0.13 ±0.04 | 0.02 ±0.01 | 95.99 | 4.0 | 467 ±97 | D.L. | 70.07 ±2 | - |
| | L int (4) | 60.3 ±1.7 | 0.67 ±0.16 | 12.88 ±1.65 | 0.09 ±0.04 | 9.57 ±1.3 | 0.19 ±0.05 | 1.78 ±0.24 | 7.16 ±0.56 | D.L. | 1.76 ±0.61 | 0.21 ±0.11 | 0.06 ±0.03 | 0.02 | 94.66 | 5.3 | 595 ±278 | D.L. | 24.95 ±2.03 | - |
| | Ol (11) | 42.1 ±1.1 | D.L. | 0.2 ±0.2 | 0.22 ±0.17 | 6.89 ±3.69 | 0.12 ±0.04 | 49.33 ±2.49 | 0.13 ±0.08 | 0.46 ±0.08 | D.L. | D.L. | D.L. | - | 99.46 | - | 1477 ±1189 | 3607 ±666 | 92.71 ±3.97 | - |
| | Opx (13) | 57.8 ±1.5 | D.L. | 0.81 ±0.34 | 0.26 ±0.12 | 2.7 ±0.28 | D.L. ±0.03 | 37.2 ±1.13 | 0.2 ±0.13 | 0.17 ±0.07 | 0.13 ±0.11 | D.L. | D.L. | - | 99.28 | - | 1791 ±789 | 1313 ±536 | 96.6 ±0.97 | - |
| | ChrMgt (2) | D.L. | 0.36 ±0.18 | 1.02 ±1.07 | 0.6 ±0.77 | 84.67 ±2.96 | 0.19 ±0.03 | 4.82 ±1.39 | D.L. | 0.65 ±0.05 | D.L. | D.L. | D.L. | - | 92.3 | - | 4071 ±5283 | 5116 ±400 | 26.95 ±7.06 | 20.7 ±13.34 |
| **P18** | L bas (122) | 49.1 ±4.3 | 1.21 ±0.12 | 14.12 ±1.31 | 0.06 ±0.01 | 7.4 ±0.68 | 0.14 ±0.04 | 12.17 ±1.33 | 7.97 ±0.7 | D.L. | 2.59 ±0.28 | 0.28 ±0.05 | 0.12 ±0.03 | - | 95.1 | 4.9 | 417 ±101 | D.L. | 74.31 ±2.56 | - |
| | L int (5) | 49.9 ±1.7 | 1.43 ±0.15 | 16.07 ±1.82 | 0.05 ±0.02 | 6.22 ±0.44 | 0.13 ±0.04 | 7.38 ±3.6 | 9.05 ±0.81 | D.L. | 3.08 ±0.35 | 0.36 ±0.06 | 0.13 ±0.03 | - | 93.8 | 6.2 | 330 ±111 | D.L. | 64.8 ±11.65 | - |
| | Ol (3) | 42.6 ±1.4 | D.L. | 0.08 ±0.01 | 0.09 ±0.02 | 6.28 ±0.14 | 0.11 ±0.02 | 49.32 ±0.41 | 0.15 ±0.01 | 0.38 ±0.02 | D.L. | D.L. | D.L. | - | 99.0 | - | 643 ±151 | 2960 ±161 | 93.34 ±0.09 | - |
| | Opx (4) | 56.3 ±1.4 | D.L. | 0.9 ±0.6 | 0.25 ±0.03 | 3.88 ±0.33 | 0.1 ±0.02 | 37.09 ±1.4 | 0.66 ±0.34 | 0.14 ±0.03 | D.L. | D.L. | D.L. | - | 99.3 | - | 1728 ±221 | 1084 ±272 | 97.14 ±1.03 | - |
| | Mgt (2) | 0.2 | 0.73 ±0.09 | 7.17 ±0.09 | 0.78 ±0.2 | 72.16 ±0.01 | D.L. | 9.19 ±0.06 | D.L. | 0.65 ±0.02 | D.L. | D.L. | D.L. | - | 90.9 | - | 5323 ±1374 | 5073 ±139 | 48.31 ±0.46 | 6.78 ±1.72 |
| **P36** | L bas (21) | 50.6 ±0.3 | 1.36 ±0.03 | 13.45 ±0.13 | 0.07 | 8.07 ±0.13 | 0.16 ±0.03 | 11.5 ±0.29 | 9.16 ±0.14 | 0.02 ±0.01 | 2.89 ±0.14 | 0.29 ±0.05 | 0.14 ±0.02 | - | 97.72 | 2.2 | 513 ±22 | D.L. | 71.73 ±0.74 | - |
| **P3** | L bas (151) | 49.6 ±7.2 | 1.34 ±0.21 | 13.47 ±2.11 | 0.06 ±0.01 | 8.03 ±1.17 | 0.16 ±0.05 | 10.4 ±1.78 | 9.27 ±1.37 | D.L. | 2.66 ±0.45 | 0.28 ±0.06 | 0.14 ±0.04 | 0.02 ±0.01 | 95.48 | 4.5 | 405 ±100 | D.L. | 68.65 ±7.59 | - |
| | L int (16) | 49.9 ±1.5 | 1.51 ±0.27 | 14.2 ±1.78 | 0.08 ±0.03 | 8.74 ±0.75 | 0.17 ±0.05 | 9.95 ±4.82 | 10.36 ±1.68 | D.L. | 2.44 ±0.31 | 0.27 ±0.07 | 0.15 ±0.04 | - | 97.72 | 2.3 | 516 ±234 | D.L. | 64.76 ±7.32 | - |
| | Ol (17) | 43.2 ±1.4 | D.L. | 0.97 ±0.83 | 0.07 ±0.03 | 7.39 ±1.42 | 0.15 ±0.03 | 45.91 ±2.51 | 0.81 ±0.7 | 0.36 ±0.05 | D.L. | D.L. | D.L. | - | 98.84 | - | 508 ±181 | 2522 ±1013 | 91.72 ±1.53 | - |
| | Opx (11) | 54.2 ±0.8 | 0.19 ±0.13 | 3.08 ±1.53 | 0.24 ±0.06 | 4.48 ±0.54 | 0.14 ±0.03 | 34.3 ±2.84 | 1.7 ±0.86 | 0.16 ±0.03 | 0.41 ±0.31 | 0.05 ±0.03 | D.L. | - | 98.95 | - | 1609 ±390 | 1238 ±264 | 96.09 ±2.38 | - |
| | ChrMgt | 0.6 ±0.1 | 2.24 ±0.14 | 17.9 ±0.01 | 2.69 ±0.27 | 57.52 ±0.6 | 0.15 | 11.72 ±0.32 | D.L. | - | D.L. | D.L. | D.L. | - | 92.84 | - | 18371 ±1858 | - | 53.55 ±1.35 | 9.14 ±0.85 |
| **P7** | L bas (75) | 50.6 ±5.9 | 1.22 ±0.15 | 12.78 ±1.48 | 0.07 ±0.01 | 8.02 ±0.93 | 0.16 ±0.03 | 12.96 ±1.52 | 8.78 ±1.03 | D.L. | 2.49 ±0.29 | 0.27 ±0.04 | 0.16 ±0.03 | - | 97.54 | 2.5 | 508 ±62 | D.L. | 73.68 ±4.84 | |

[a] The experiment number, conditions are given in the Table 2. [b] The produced mineral and melt phases. Bracketed numbers correspond to the number of analyses. $L_{bas}$, and $L_{int}$ are compositions of the basaltic and interstitial glasses correspondingly; Ol – olivine, Opx - orthopyroxene, Chr – chromite, ChrMgt - chromiferous magnetite, Mgt - magnetite. [c] All iron content is recalculated as iron total ($FeO_{tot}$). [d] Magnesium number [100Mg/(Mg+$Fe^{2+}$)] in atoms per formula unit. In case of glasses iron total is taken [e] Chromium number [100Cr/(Cr+Al)] in atoms per formula unit. D.L. corresponds to values below detection limit.

**Table 4. Average major and trace element composition of the 0.5 - 1.0 GPa homogeneous samples obtained by LA-ICP-MS**

| Sample | $SiO_2$, wt% | $TiO_2$ | $Al_2O_3$ | FeO | MgO | MnO | CaO | $Na_2O$ | $K_2O$ | $P_2O_5$ |
|---|---|---|---|---|---|---|---|---|---|---|
| P7 (47) [a] | 51.14 ± 0.37 | 1.24 ± 0.02 | 12.63 ± 0.23 | 8.97 ± 0.15 | 12.64 ± 0.18 | 0.16 ± 0.01 | 9.09 ± 0.15 | 2.67 ± 0.05 | 0.26 ± 0.01 | 0.22 ± 0.01 |
| P36 (43) | 51.90 ± 0.49 | 1.26 ± 0.02 | 13.49 ± 0.38 | 8.87 ± 0.17 | 10.85 ± 0.59 | 0.16 ± 0.01 | 9.31 ± 0.22 | 2.68 ± 0.07 | 0.27 ± 0.01 | 0.20 ± 0.01 |
| | Li, ppm | Cr | Ni | Rb | Sr | Y | Zr | Nb | Ba | La |
| P7 (47) | 5.92 ± 0.29 | 505.9 ±10.07 | 204.6 ± 29.37 | 6.99 ± 0.22 | 138.03 ± 3.26 | 24.84 ± 0.59 | 8.74 ± 2.15 | 8.21 ±0.20 | 75.38 ± 2.12 | 6.68 ± 0.20 |
| P36 (43) | 16.79 ± 0.77 | 511.6 ± 38.6 | 136.9 ± 30.2 | 7.09 ± 0.22 | 141.2 ± 2.62 | 25.8 ± 0.86 | 92.94±2.58 | 8.85± 0.30 | 76.93 ±1.70 | 6.64 ± 0.20 |
| ppm | Ce | Pr | Nd | Sm | Eu | Gd | Tb | Dy | Ho | Er |
| P7 (47) | 16.36 ± 0.33 | 2.36 ± 0.08 | 11.43 ± 1.1 | 3.39 ± 0.23 | 1.19 ± 0.07 | 4.04 ± 0.21 | 0.68 ± 0.04 | 4.40 ± 0.24 | 0.93 ± 0.06 | 2.67 ± 0.20 |
| P36 (43) | 16.63 ± 0.38 | 2.42 ± 0.08 | 11.51 ± 0.37 | 3.48 ± 0.16 | 1.23 ± 0.07 | 4.24 ± 0.21 | 0.71 ± 0.04 | 4.72 ± 0.20 | 0.98 ± 0.05 | 2.88 ± 0.19 |
| ppm | Tm | Yb | Lu | Hf | Th | U | $(La/Sm)_n$ | $(La/Yb)_n$ | | |
| P7 (47) | 0.39 ± 0.03 | 2.57 ± 0.16 | 0.39 ± 0.03 | 2.38 ± 0.10 | 0.60 ± 0.03 | 0.21 ± 0.02 | 1.28 ± 0.08 | 1.87 ± 0.11 | | |
| P36 (43) | 0.40 ± 0.03 | 2.67 ± 0.14 | 0.41 ± 0.03 | 2.53 ± 0.11 | 0.63 ± 0.04 | 0.21 ± 0.02 | 1.23 ± 0.06 | 1.79 ± 0.10 | | |

[a] The experiment number, conditions are given in the Table 2. Bracketed numbers correspond to the number of analyses. The average composition and the glass homogeneity is represented as 1 σ std. deviation. La/Sm and La/Yb are normalized to the composition of the primitive mantle (Lyubetskaya and Korenaga, 2007).

**Table 5. Calculation of assimilation rate at 0.5 - 1.0 GPa and 1300°C**

| № | Pressure (GPa) | Additional water (wt.%) | Duration (s) | Run duration ($s^{1/2}$) | Volume difference* ($mm^3$) | Volume difference (%) | Linear difference (μm) | Linear rate1 ($\mu m/s^{1/2}$) | Linear rate2 ($\mu m^2/s$) | Relative rate ($m^2/s$) | Ln(Average rate) ($m^2/s$) |
|---|---|---|---|---|---|---|---|---|---|---|---|
| P15 | 0.5 | - | 1800 | 42.43 | 2.210 | 53.9 | 1302.6 | 30.7 | 9.43E+02 | 9.43E-10 | |
| P10 | 0.5 | - | 9300 | 96.44 | 2.834 | 69.2 | 1415.1 | 14.7 | 2.15E+02 | 2.15E-10 | |
| P18 | 0.5 | - | 18000 | 134.16 | 3.602 | 85.1 | 1532.9 | 11.4 | 1.31E+02 | 1.31E-10 | |
| Avr. | 0.5 | - | - | - | - | - | - | - | - | 4.30E-10 | -21.6 |
| P20 | 0.5 | 7.8 | 1800 | 42.43 | 1.830 | 43.3 | 1223.1 | 28.8 | 8.31E+02 | 8.31E-10 | |
| P21 | 0.5 | 9.5 | 9000 | 94.87 | 3.295 | 80.6 | 1488.0 | 15.7 | 2.46E+02 | 2.46E-10 | |
| P26 | 0.5 | 6.4 | 18000 | 134.16 | 2.195 | 51.9 | 1299.7 | 9.7 | 9.38E+01 | 9.38E-11 | |
| Avr. | 0.5 | - | - | - | - | - | - | - | - | 3.90E-10 | -21.7 |
| P3 | 1.0 | - | 9000 | 94.87 | 1.318 | 32.2 | 1096.5 | 11.6 | 1.34E+02 | 1.34E-10 | -22.7 |

* volume calculated as cylinder volume based on initial 3D measurements of the serpentinite disc shape and based on measurements of 2D dimensions of the residual serpentinite using SEM in the polished sections. The volume difference is calculated as $(V_{int} – V_{exp})/V_{int}$, where $V_{exp}$ is volume after experimentation and $V_{int}$ is initial volume. Volumes have been calculated from the serpentinite mass and density for the initial volume ($V_{int}$) and the 2D size measured using SEM after experimentation ($V_{exp}$). The uncertainty on the assimilation rate (relative rate) is in the limit of 15 rel. %. Avr. – average rate values.

**Supplementary Table A1**

**Composition of mineral and glass phases from the 0.5 - 1.0 GPa basaltic melt-serpentinite reaction experiments (P10, P12, P20, P21, P26)**

| Exp [a] | Phase [b] | $SiO_2$ (wt%) | $TiO_2$ (wt%) | $Al_2O_3$ (wt%) | $Cr_2O_3$ (wt%) | $FeO_{tot}$ [c] (wt%) | MnO (wt%) | MgO (wt%) | CaO (wt%) | NiO (wt%) | $Na_2O$ (wt%) | $K_2O$ (wt%) | $P_2O_5$ (wt%) | Cl (wt%) | Total (wt%) | $H_2O$ (wt%) | Cr (ppm) | Ni (ppm) | Mg# [d] |
|---|---|---|---|---|---|---|---|---|---|---|---|---|---|---|---|---|---|---|---|
| **P10** | L bas (33) | 52.7 ±3.0 | 1.45 ±0.13 | 16.38 ±1.68 | 0.03 ±0.02 | 7.2 ±1.16 | 0.13 ±0.04 | 4.74 ±2.45 | 8.73 ±1.60 | D.L. | 2.54 ±0.66 | 0.35 ±0.06 | 0.17 ±0.02 | 0.6 ±0.48 | 95.4 | 4.6 | 209 ±147 | D.L. | 50.84 ±9.09 |
| | L int (56) | 54.3 ±1.3 | 1.48 ±0.16 | 15.31 ±0.93 | D.L. | 5.83 ±0.57 | 0.13 ±0.03 | 4.97 ±1.99 | 9.95 ±0.87 | D.L. | 2.82 ±1.34 | 0.36 ±0.04 | 0.17 ±0.03 | 0.04 ±0.01 | 95.4 | 4.6 | D.L. | D.L. | 58.9 ±6.5 |
| | Ol (49) | 40.9 ±0.6 | D.L. | D.L. | 0.11 ±0.25 | 6.97 ±0.88 | 0.14 ±0.04 | 48.88 ±2.35 | 0.34 ±0.35 | 0.33 ±0.07 | D.L. | D.L. | D.L. | - | 98.5 | - | 488 ±106 | 2582 ±559 | 92.6 ±1.2 |
| | Cpx (1) | 48.3 | D.L. | 7.3 | D.L. | 6.1 | 0.15 | 14.9 | 20.13 | 0.06 | 0.44 | D.L. | D.L. | - | 99.0 | - | D.L. | 456 | 81.3 |
| **P12** | L bas (177) | 49.44 ±3.86 | 1.22 ±0.11 | 12.65 ±1.01 | 0.07 ±0.01 | 8.92 ±0.69 | 0.15 ±0.03 | 12.10 ±0.93 | 8.71 ±0.67 | 0.03 ±0.03 | 2.57 ±0.21 | 0.34 ±0.04 | 0.16 ±0.03 | - | 96.29 | 3.7 | 511 ±85 | D.L. | 70.75 ±0.40 |
| **P20** | L bas (86) | 49.60 ±1.44 | 1.22 ±0.15 | 13.01 ±1.17 | 0.07 ±0.02 | 8.33 ±0.28 | 0.15 ±0.04 | 11.03 ±2.46 | 8.48 ±1.17 | 0.04 ±0.02 | 2.27 ±0.20 | 0.27 ±0.04 | 0.15 ±0.03 | 0.02 ±0.01 | 94.54 | 5.5 | 493 ±161 | D.L. | 69.49 ±5.0 |
| | L int (11) | 57.28 ±2.75 | 1.06 ±0.29 | 13.69 ±2.12 | 0.06 ±0.03 | 5.78 ±0.77 | 0.10 ±0.04 | 3.79 ±0.91 | 9.06 ±1.04 | D.L. | 2.55 ±0.50 | 0.34 ±0.04 | N.A. | N.A. | 93.76 | 6.2 | 435 ±164 | D.L. | 51.13 ±12.21 |
| | Ol (27) | 41.44 ±0.44 | D.L. | D.L. | 0.09 ±0.02 | 7.01 ±0.68 | 0.11 ±0.04 | 49.81 ±0.71 | 0.13 ±0.06 | 0.44 ±0.07 | D.L. | D.L. | D.L. | - | 99.11 | - | 608 ±133 | 3455 ±588 | 92.70 ±0.74 |
| | Opx (1) | 58.32 | D.L. | 0.87 | 0.30 | 2.85 | 0.15 | 36.74 | 0.00 | 0.17 | 0.03 | D.L. | D.L. | - | 99.4 | - | 2018 | 1297 | 95.83 |
| **P21** | L bas (107) | 49.32 ±1.72 | 1.22 ±0.06 | 13.65 ±0.85 | 0.07 ±0.01 | 7.81 ±0.62 | 0.15 ±0.04 | 11.44 ±2.08 | 8.57 ±0.79 | D.L. | 2.42 ±0.18 | 0.26 ±0.03 | 0.16 ±0.02 | - | 94.98 | 5.0 | 480 ±95 | D.L. | 72.0 ±3.7 |
| | L int (9) | 57.55 ±1.51 | 1.19 ±0.09 | 14.17 ±0.84 | 0.11 ±0.03 | 5.23 ±0.39 | 0.13 ±0.03 | 6.23 ±1.97 | 8.56 ±0.50 | D.L. | 2.75 ±0.16 | 0.36 ±0.07 | 0.08 | N.A. | 96.3 | 3.7 | 733 ±232 | D.L. | 66.7 ±6.6 |
| | Ol (11) | 41.48 ±0.63 | D.L. | D.L. | 0.08 ±0.02 | 5.42 ±0.38 | 0.11 ±0.03 | 49.76 ±0.95 | 0.14 ±0.02 | 0.33 ±0.07 | D.L. | D.L. | D.L. | - | 99.0 | - | 535 ±127 | 2565 ±553 | 94.23 ±0.47 |
| | Opx (9) | 57.5 ±1.95 | D.L. | D.L. | 0.41 ±0.34 | 2.97 ±0.46 | 0.12 ±0.02 | 34.5 ±3.9 | 0.82 ±0.63 | 0.14 ±0.02 | D.L. | D.L. | D.L. | - | 99.5 | - | 2831 ±2348 | 1135 ±164 | 95.3 ±0.9 |
| **P26** | L bas (85) | 49.94 ±7.78 | 1.26 ±0.2 | 13.35 ±2.2 | 0.07 ±0.02 | 7.32 ±1.13 | 0.16 ±0.04 | 10.30 ±1.72 | 9.21 ±1.44 | D.L. | 2.66 ±0.42 | 0.29 ±0.05 | 0.16 ±0.02 | - | 94.6 | 5.4 | 495 ±158 | D.L. | 70.68 ±5.48 |
| | L int (9) | 54.43 ±0.78 | 1.36 ±0.1 | 14.35 ±0.64 | 0.09 ±0.01 | 6.64 ±0.32 | 0.15 ±0.04 | 5.25 ±0.87 | 11.17 ±0.59 | D.L. | 2.97 ±0.16 | 0.33 ±0.03 | 0.13 | - | 96.8 | 3.2 | 600 ±54 | D.L. | 58.2 ±3.8 |
| | Ol (20) | 41.00 ±0.29 | D.L. | D.L. | 0.06 ±0.01 | 8.35 ±0.17 | 0.16 ±0.04 | 48.00 ±0.53 | 0.25 ±0.08 | 0.21 ±0.03 | D.L. | D.L. | D.L. | - | 99.16 | - | 421 ±90 | 1649 ±255 | 91.11 ±0.21 |
| | Opx (8) | 57.1 ±0.53 | 0.10 ±0.05 | 0.97 ±0.34 | 0.35 ±0.10 | 4.83 ±0.53 | 0.13 ±0.03 | 33.53 ±1.04 | 1.5 ±0.3 | 0.11 ±0.04 | D.L. | D.L. | D.L. | - | 99.7 | - | 2425 ±663 | 872 ±289 | 92.5 ±0.9 |

[a] The experiment number, conditions are given in the Table 2. [b] The produced mineral and melt phases. Bracketed numbers correspond to the number of analyses. $L_{bas}$, and $L_{int}$ are compositions of the basaltic and interstitial glasses correspondingly; Ol – olivine, Opx - orthopyroxene, Cpx - clinopyroxene. [c] All iron content is recalculated as iron total ($FeO_{tot}$). [d] Magnesium number [100Mg/(Mg+$Fe^{2+}$)] in atoms per formula unit. In case of glasses, total iron is taken. D.L. corresponds to values below detection limit; N.A. – not analyzed.